\documentclass[12pt]{emulateapj}
\usepackage{apjfonts}
\usepackage{natbib}

\shorttitle{Dust in Early-Type Galaxies} 
\shortauthors{Martini, Dicken, \& Storchi-Bergmann} 
\slugcomment{ApJ Accepted [16 February 2013]} 

\newcommand{\spitzer}{{\it Spitzer}} 
\newcommand{\herschel}{{\it Herschel}}

\newcommand{\hst}{{\it HST}} 
\newcommand{\msun}{M$_\odot$}
\newcommand{\msunyr}{M$_\odot$ yr$^{-1}$}
\newcommand{\ntot}{38} 		
\newcommand{\nmips}{eight} 	

\begin{document}

\title{The Origin of Dust in Early-Type Galaxies and Implications for Accretion onto Supermassive Black Holes} 

\author{Paul Martini\altaffilmark{1}}
\affil{Department of Astronomy Department and Center for Cosmology and Astroparticle Physics, The Ohio State University, Columbus, OH 43210}
\altaffiltext{1}{Visiting Astronomer, North American ALMA Science Center and University of Virginia, Charlottesville, VA} 

\author{Daniel Dicken} 
\affil{Institut de Astrophysique Spatiale, Paris, France} 

\and 

\author{Thaisa Storchi-Bergmann}
\affil{Instituto de F\'isica, Universidade Federal do Rio Grande do Sul, Av. Bento Gon{\c c}alves 9500, Caixa Postal 15051, 91501-970 Porto Alegre, RS Brasil}

\begin{abstract}

We have conducted an archival \spitzer\ study of \ntot\ early-type galaxies in order
to determine the origin of the dust in approximately half of this population.
Our sample galaxies generally have good wavelength coverage from $3.6\mu$m to
$160\mu$m, as well as visible-wavelength \hst\ images. We use the \spitzer\
data to estimate dust masses, or establish upper limits, and find that all of
the early-type galaxies with dust lanes in the \hst\ data are detected in all 
of the \spitzer\ bands and have dust masses of $\sim 10^{5-6.5} M_\odot$, while 
galaxies without dust lanes are not detected at $70\mu$m and $160\mu$m and 
typically have
$<10^5 M_\odot$ of dust. The apparently dust-free galaxies do have $24\mu$m
emission that scales with the shorter wavelength flux, yet substantially
exceeds the expectations of photospheric emission by approximately a factor 
of three. We conclude this
emission is dominated by hot, circumstellar dust around evolved
stars that does not survive to form a substantial interstellar component. The
order of magnitude variations in dust masses between galaxies with similar
stellar populations rules out a subtantial contribution from continual,
internal production in spite of the clear evidence for circumstellar dust.
We demonstrate that the interstellar dust is not due to purely 
external accretion, unless the product of the merger rate of dusty 
satellites and the dust lifetime is at least an order of magnitude higher 
than expected. We propose that dust in early-type galaxies is seeded by 
external accretion, yet the accreted dust is maintained by continued 
growth in externally-accreted cold gas beyond the nominal lifetime of 
individual grains. The several Gyr depletion time of the cold gas is long 
enough to reconcile the fraction of dusty early-type galaxies with the merger 
rate of gas-rich satellites. As the majority of dusty early-type galaxies 
are also low-luminosity Active Galactic Nuclei and likely fueled by this cold 
gas, their lifetime should similarly be several Gyr.

\end{abstract}


\section{Introduction} \label{sec:intro} 


One of the most striking characteristics of elliptical and lenticular galaxies 
is their apparent uniformity. Much more than their later-type cousins, 
early-type galaxies appear to have quite smooth and symmetric surface 
brightness profiles. Their morphological self-similarity is largely because 
most early-type galaxies, and in particular ellipticals, are expected to 
be the result of nearly equal-mass mergers that have produced symmetric 
stellar distributions via dynamical relaxation. This assembly process, 
combined with the generally large halos that host early-type galaxies, has 
also substantially heated the gas in their interstellar medium (ISM) such 
that cooling and subsequent star formation are quite inefficient. As a result, 
their stellar populations tend to be mostly old and well-mixed, which 
reinforces their similar and smooth appearance. 

Against the backdrop of these similarities, and the simplicity of this general 
picture, the differences between nominally similar early-type galaxies are cast 
into sharper relief than for the more highly structured late-type disk galaxy 
population. One striking difference is that some, but not all, early-type 
galaxies clearly contain cold, interstellar dust. This was revealed in IRAS 
observations that detected a number of elliptical galaxies at $100\mu$m 
\citep{jura87,knapp89,goudfrooij95,bregman98}. A study by \citet{knapp92} also 
demonstrated that ellipticals had emission at $10-12\mu$m in excess of what 
was expected from stellar photospheres alone. This excess could be explained 
by circumstellar dust emission from evolved stars and proved to be a strong 
observational confirmation of expectations for mass loss from old stellar 
populations.

The presence of dust in early-type galaxies is interesting because it may 
provide additional information about their formation histories, the properties 
of dust when embedded within a substantially greater fraction of hot gas than 
is present in typical spirals, and the evolution of the interstellar medium, 
in addition to stellar mass loss from evolved stars. Yet the presence of dust 
is also in apparent conflict with the expectation that the dust destruction 
timescale should be relatively short ($\tau_{dust} \sim 2 \times 10^{4}$ yr), 
largely due to sputtering of small grains in their hot ISM \citep{draine79a}. 

Observations with \hst\ revealed that dust lanes were common and probably 
present in the majority of early-type galaxies. These observations showed dust 
lanes in 
absorption within 100s of parsecs to a kpc of the nucleus \citep{vandokkum95}. 
This study also noted that the incidence of dust is higher is radio galaxies 
than in radio-quiet galaxies, estimated that the mass of dust ranged from 
$10^{3-7}$ \msun, and that the distribution of the dust lanes suggests external 
accretion. Subsequent studies of atomic \citep{morganti06} and molecular gas 
kinematics \citep{young11,davis11} supported external accretion because the 
gas kinematics are often decoupled from the stellar kinematics. 

While there is good, anecdotal evidence for external accretion, the evolved 
stars in early-type galaxies are also a potential source for the dust 
\citep{knapp92}. \citet{athey02} showed that the mid-infrared emission from 
early-type galaxies is consistent with stellar mass loss of on order $0.1 - 
1$ \msunyr\ for a typical $L*$ galaxy. For a dust-to-gas ratio of 
$\epsilon_{dg} = 0.005$ and a dust lifetime of $10^{7-8}$ yr characteristic of 
the Milky Way, this would lead to a typical reservoir of $5 \times 10^{3-5}$ 
\msun\ (or $10-100$ \msun\ for $\tau_{dust} = 2 \times 10^{4}$ yr). While this 
dust would initially match the spatial and kinematic distribution of the 
evolved stellar population, \citet{mathews03} suggest that cooling in 
dust-rich gas could lead to 
clumping and settling into the central kiloparsec on a timescale of $10^{8-9}$ 
yr. While this substantially exceeds the expected lifetime of individual dust 
grains, the clumpiness may sufficiently shield the dust to produce a steady 
state dust mass that is consistent with the observations. However, the 
internal origin hypothesis does not explain the dramatic variation in dust 
mass between otherwise quite similar galaxies. \citet{temi04} showed that there 
is no correlation between the far-infrared luminosity of early-type galaxies 
and their visible-wavelength luminosities, even though visible-wavelength 
light should correlate reasonably well with the number of mass-losing stars. 
This absence of a correlation can be more readily explained by the external 
accretion hypothesis. 

Another important aspect of these studies was the observation that the 
majority of the early-type galaxies with dust also have emission-line nuclei 
that often indicate the presence of a LINER or Seyfert galaxy 
\citep{vandokkum95,ravindranath01,lauer05}. In order to investigate this 
correlation with a well-selected sample of early-type galaxies, in 
\citet{lopes07} we studied archival \hst\ images of a carefully-matched 
sample of active and inactive early-type hosts and determined that 
all active early-type galaxies also have dust within 100s of parsecs of their 
nuclei, while only $\sim 25$\% of inactive galaxies had evidence for dust. 
This study showed that $\sim 60$\% of the early-type galaxy population has 
interstellar dust. As we noted in \citet{lopes07}, the high incidence of dust 
in early-type galaxies presents a key challenge to the external origin 
hypothesis, namely that $\sim 60$\% of the early-type galaxy population must 
have accreted a gas and dust-rich dwarf within a time period comparable to 
the dust destruction time. For a dust destruction timescale of $10^{7-8}$ yr, 
this implies a very high rate of gas-rich mergers. 

A missing ingredient in the \citet{lopes07} study is that there was 
no constraint on the dust mass distribution in the early-type galaxies, which 
would provide a useful constraint on the mass range and consequently 
mass ratio of the satellite galaxies required to provide the observed 
mass of dust. \citet{lopes07} also employed the structure map technique 
of \citet{pogge02} to identify dust lanes. As this is a contrast 
enhancement technique, it is insensitive to uniform or diffusely 
distributed dust. Uniformly-distributed dust in the galaxies without 
dust lanes in the central kpc would provide strong support for the internal 
origin hypothesis. 

Over the last few years, several studies with \spitzer\ and \herschel\ have 
assembled far-infrared measurements and estimated dust masses and basic 
dust properties for many early-type galaxies. One result of this work 
is that the dust in early-type galaxies is warmer than in spiral galaxies 
\citep{bendo03,skibba11,smith12,auld13}, which may be due to more intense 
radiation fields or different dust grain properties. The \herschel\ KINGFISH 
survey showed that early-type galaxies exhibit a strong correlation between 
the dust to stellar flux ratio and the specific star formation rate, which 
could be due to low levels of on-going star formation \citep{skibba11}. 
The \herschel\ Reference Survey \citep{smith12} and the \herschel\ 
Virgo Cluster Survey \citep{auld13} detect many early-type galaxies 
and measure dust masses in the range of $\sim 10^{5-7}$ \msun. 
The \citet{smith12} study concluded that much of this dust was acquired from 
interactions due to the wide range in dust to stellar mass ratio. A similar 
conclusion was reached by \citet{rowlands12}, who detected 5.5\% of luminous 
ETGs in the \herschel-ATLAS/GAMA study. While only sensitive to quite high 
dust masses (their mean detected dust mass was $5.5 \times 10^7$ \msun), they 
conclude that these high dust masses could not be produced internally. 

In this paper we present a \spitzer\ archival study of the origin of the 
dust in early-type galaxies. Our sample is composed of galaxies in the 
\citet{lopes07} early-type galaxy sample that have both IRAC and MIPS 
observations. We describe this sample in 
\S\ref{sec:sample} and our data processing and photometry in \S\ref{sec:data}. 
All of the galaxies in this sample are detected in the IRAC bands, which 
are dominated by stellar emission, as well as in the $24\mu$m MIPS band. 
Many are also detected in the longer wavelength MIPS bands. The $24\mu$m 
emission from galaxies not detected at longer wavelengths appears to be 
dominated by circumstellar dust. We describe how we account for this 
emission in \S\ref{sec:stardust}. To derive the dust masses and 
properties of the dust, we fit these galaxies with the dust models of 
\citet{draine07a}. These fits and the derived properties are described 
in \S\ref{sec:dust}. We use these data to develop a new hypothesis for the 
origin of the dust in early-type galaxies in \S\ref{sec:origin} and 
summarize our results in \S\ref{sec:summary}. 




\section{Sample} \label{sec:sample} 

\begin{figure}
\plotone{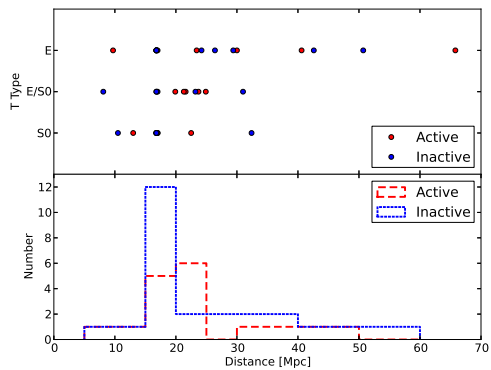}
\caption{
Sample distribution as a function of morphological type, activity, 
and distance. ({\it Top panel}) Active and inactive galaxies 
({\it red} and {\it blue} circles, respectively) as a function of 
morphological type and distance. ({\it Bottom panel}) Histogram 
of all morphological types of the active and inactive galaxies 
({\it red} and {\it blue} lines, respectively) as a function of
distance. The substantial overdensity at 16.8 Mpc is because many 
galaxies in that bin are members of the Virgo cluster. 
\label{fig:sample} 
}
\end{figure}

Our sample is composed of \ntot\ bright, nearby, early-type galaxies (types E, 
E/S0, and S0). They constitute all of the early-type galaxies from the \hst\ 
study of \citet{lopes07} that have observations in at least two MIPS bands. 
The goal of the \citet{lopes07} study was to compare dust structure in the 
central 100s of parsecs between active and inactive galaxies and that sample 
contained equal numbers of well-matched active and inactive galaxies from the 
the Palomar survey of \citet{ho95}. The Palomar survey is very well suited for 
this purpose because it consists of very high signal-to-noise ratio (SNR) 
spectroscopy of all bright galaxies in a fixed region of the sky. Furthermore, 
all of the nuclear spectra have been classified as either absorption-line 
nuclei, HII galaxies, LINERs, transition objects, or Seyfert galaxies 
\citep{ho97a}. The Palomar survey sample is thus both unbiased with respect to 
nuclear activity and has homogeneous and sensitive nuclear classifications. 
A further advantage is that there are a large number of ancillary data 
products, which are described in \citet{ho08}. 

The \citet{lopes07} study began with all of the LINERs and Seyfert galaxies 
(hereafter simply active galaxies) in the Palomar sample with \hst\ 
observations and, for each active galaxy, identified a control (inactive) 
galaxy with the same morphology, distance, luminosity, and axis ratio. They 
identified 26 early-type pairs with relatively strict matching criteria 
(morphological $T$ type within 1 unit, absolute $B$ magnitude within 1 mag, 
distance within 50\%) and eight additional pairs of early-type galaxies that 
were not as well matched. These 34 pairs of active and inactive early-type 
galaxies had very similar distributions of distances, luminosities, and 
morphological types and are consequently very well suited to compare the 
differences between active and inactive early-type galaxies. As noted in the 
Introduction, that \hst\ study found that all active, early-type galaxies 
possess some form of dust structure within several 100 pc of their nuclei, 
whereas only 26\% (nine) of the inactive galaxies exhibited 
evidence for such dust. These data thus show there is a strong dichotomy 
between dusty and active early-type galaxies and typically dust-free and 
inactive 
early-type inactive galaxies. Furthermore, as \citet{ho97a} found that 50\% of 
all early-type galaxies are classified as LINERs or Seyferts (active), this 
result also indicates that about 60\% of all early-type galaxies have dust. 

We constructed our sample from all of the early-type galaxies in the 
\citet{lopes07} study that have observations in at least two MIPS bands, and 
in most cases all three are available. All of the galaxies also 
have IRAC observations, in addition to the \hst\ observations. This sample 
has approximately equal numbers of active and inactive galaxies and 
should be an unbiased subset of the \citet{lopes07} sample because these 
galaxies were targeted for MIPS observations by a multitude of small programs. 
Because of this selection, and a desire to not decrease the sample size 
further, we did not attempt to create a pair-matched sample from this subset. 
We did confirm that there are not obvious trends between the active and 
inactive subsets as a function of distance and morphological type. 
Figure~\ref{fig:sample} shows the distance distribution for galaxies 
classified as type E, E/S0, and S0. While the true ellipticals extend out to 
larger distances than the two other types, the distribution of active and 
inactive galaxies are similar. The morphological classification, activity type, 
and distance for all of the galaxies are presented in Table~\ref{tbl:sample}, 
along with whether or not they were observed to have dust structures in the 
study of \citet{lopes07}. 

\begin{deluxetable}{lrrcc}
\tablecolumns{5}
\tablewidth{0.0truein}
\tabletypesize{\scriptsize}
\tablecaption{Sample Properties\label{tbl:sample}}
\tablehead{
\colhead{Name} &
\colhead{Morph} &
\colhead{Activity} &
\colhead{Distance} &
\colhead{HST Dust?} \\ 
\colhead{(1)} &
\colhead{(2)} &
\colhead{(3)} &
\colhead{(4)} &
\colhead{(5)} 
}
\startdata
NGC0315 .................... &     E &    active &   65.8 & Y \\
NGC0821 .................... &  E/S0 &  inactive &   23.2 & N \\
NGC1023 .................... &    S0 &  inactive &   10.5 & N \\
NGC2300 .................... &  E/S0 &  inactive &   31.0 & N \\
NGC2768 .................... &  E/S0 &    active &   23.7 & Y \\
NGC2787 .................... &    S0 &    active &   13.0 & Y \\
NGC3226 .................... &     E &    active &   23.4 & Y \\
NGC3377 .................... &  E/S0 &  inactive &    8.1 & Y \\
NGC3414 .................... &    S0 &    active &   24.9 & Y \\
NGC3607 .................... &    S0 &    active &   19.9 & Y \\
NGC3640 .................... &     E &  inactive &   24.2 & N \\
NGC3945 .................... &    S0 &    active &   22.5 & Y \\
NGC3998 .................... &    S0 &    active &   21.6 & Y \\
NGC4026 .................... &    S0 &  inactive &   17.0 & Y \\
NGC4138 .................... &    S0 &    active &   17.0 & Y \\
NGC4278 .................... &     E &    active &    9.7 & Y \\
NGC4291 .................... &     E &  inactive &   29.4 & N \\
NGC4293 .................... &   S0a &    active &   17.0 & Y \\
NGC4365 .................... &     E &  inactive &   16.8 & N \\
NGC4371 .................... &    S0 &  inactive &   16.8 & Y \\
NGC4382 .................... &    S0 &  inactive &   16.8 & N \\
NGC4406 .................... &     E &  inactive &   16.8 & N \\
NGC4526 .................... &   S0a &  inactive &   16.8 & Y \\
NGC4550 .................... &    S0 &    active &   16.8 & Y \\
NGC4570 .................... &    S0 &  inactive &   16.8 & N \\
NGC4578 .................... &    S0 &  inactive &   16.8 & N \\
NGC4589 .................... &     E &    active &   30.0 & Y \\
NGC4612 .................... &    S0 &  inactive &   16.8 & N \\
NGC4621 .................... &     E &  inactive &   16.8 & N \\
NGC4636 .................... &     E &    active &   17.0 & Y \\
NGC4649 .................... &     E &  inactive &   16.8 & N \\
NGC4694 .................... &     E &  inactive &   16.8 & Y \\
NGC5077 .................... &     E &    active &   40.6 & Y \\
NGC5273 .................... &    S0 &    active &   21.3 & Y \\
NGC5308 .................... &    S0 &  inactive &   32.4 & N \\
NGC5557 .................... &     E &  inactive &   42.6 & N \\
NGC5576 .................... &     E &  inactive &   26.4 & N \\
NGC7619 .................... &     E &  inactive &   50.7 & N \\
\enddata
\tablecomments{
Properties of the galaxies in the sample. Columns are: (1) Galaxy name; 
(2) Morphological type; (3) Activity type; (4) Distance in Mpc; (5) 
Presence or absence of dust in HST data from \citet{lopes07}
References for morphology, activity, and distance are provided in 
\citet{lopes07}. 
}
\end{deluxetable}

\section{Data Processing and Photometry} \label{sec:data} 


We obtained all available IRAC and MIPS data for these galaxies from the 
\spitzer\ 
archive. The IRAC data correspond to images in bandpasses centered at 
approximately $3.6\mu$m, $4.5\mu$m, $5.7\mu$m, and $8\mu$m. For most galaxies, 
the first three bands are dominated by stellar emission, while the fourth is 
dominated by emission from polycyclic aromatic hydrocarbons (PAHs). 
For most early-type galaxies, the fourth band is also dominated by stellar 
emission, rather than PAHs, as we discuss further below. 
The MIPS data correspond to images in bandpasses with nominal centers at 
$24\mu$m, $70\mu$m, and $160\mu$m. 
Emission at $24\mu$m is typically dominated by hot dust, which is either 
interstellar or circumstellar, while emission at longer wavelengths is 
dominated by cooler, diffuse, interstelar dust. 
We typically started our analysis with the Basic Calibrated Data (BCD) for the 
MIPS observations and the Post-BCD (PBCD) products for the IRAC observations. 
In the subsections below we describe any additional processing, photometric 
measurements, and how we determined upper limits. 

\begin{figure*}
\plotone{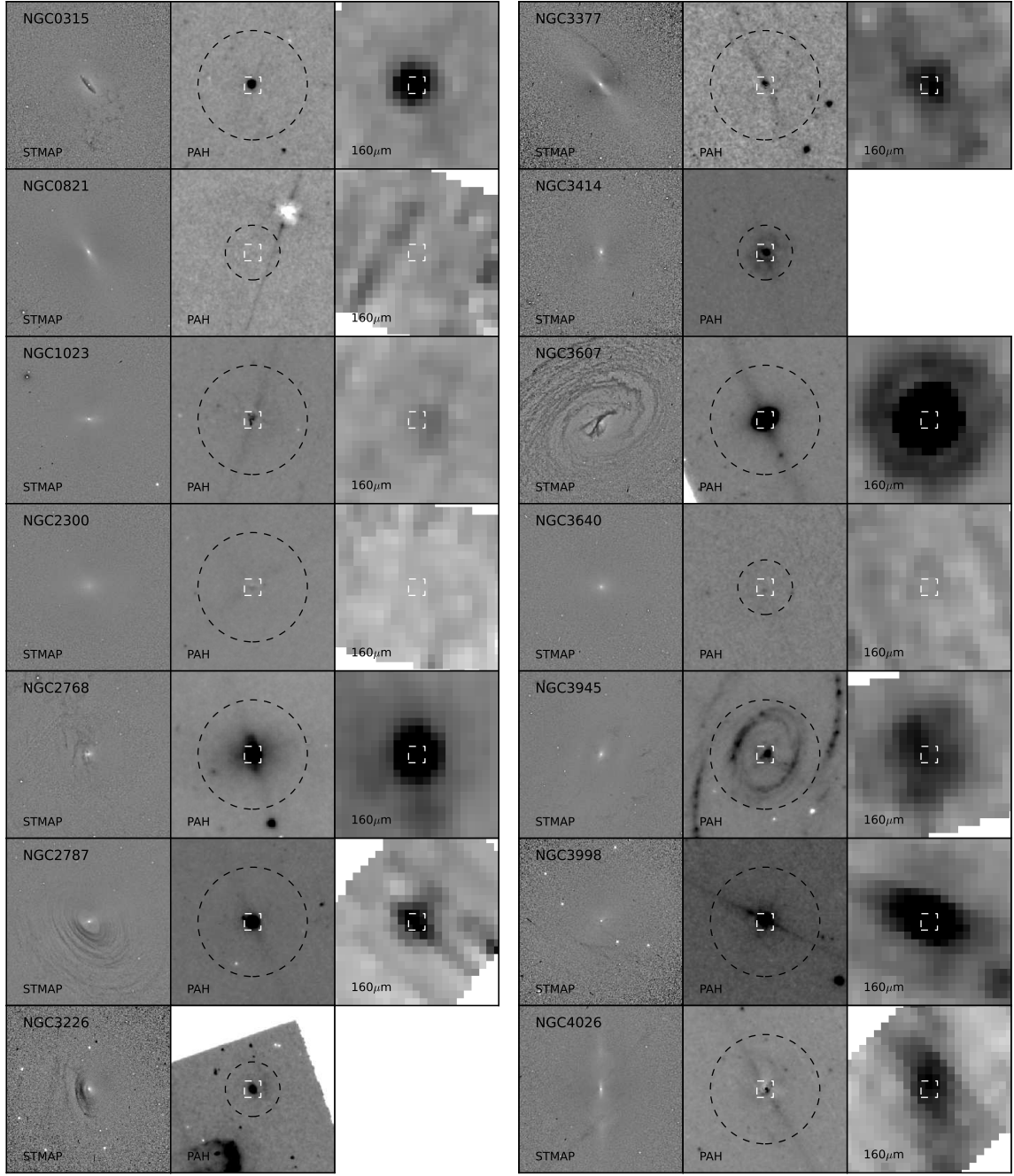} 
\caption{
Structure maps ({\it left}), PAH images ({\it middle}), and MIPS $160\mu$m 
images ({\it right}) for the sample. The structure map is 18$''$ on a 
side, while the PAH and MIPS images are 180$''$ on a side. The small 
box on the PAH and MIPS images indicates the field of view of the 
structure map. The black, dashed circle on the PAH image indicates 
the photometric aperture used for the $3.6\mu$m to $70\mu$m 
photometry (a PSF model was used for the $160\mu$m data). 
The third panel is empty if $160\mu$m MIPS data were unavailable for that 
galaxy. 
\label{fig:multi1} 
}
\end{figure*}

\begin{figure*}
\plotone{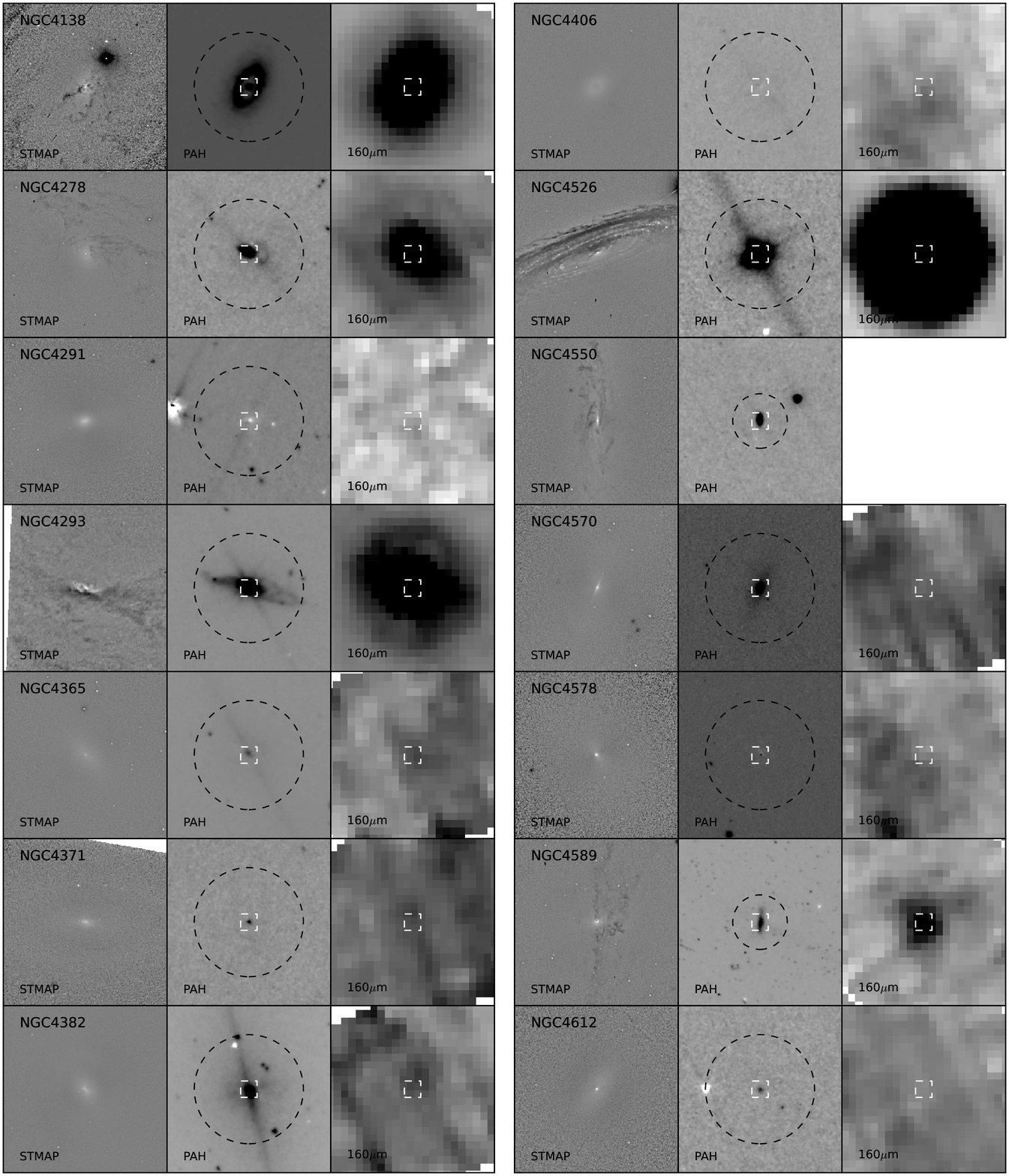} 
\caption{
Continuation of Figure~\ref{fig:multi1}
\label{fig:multi2} 
}
\end{figure*}

\begin{figure*}
\plotone{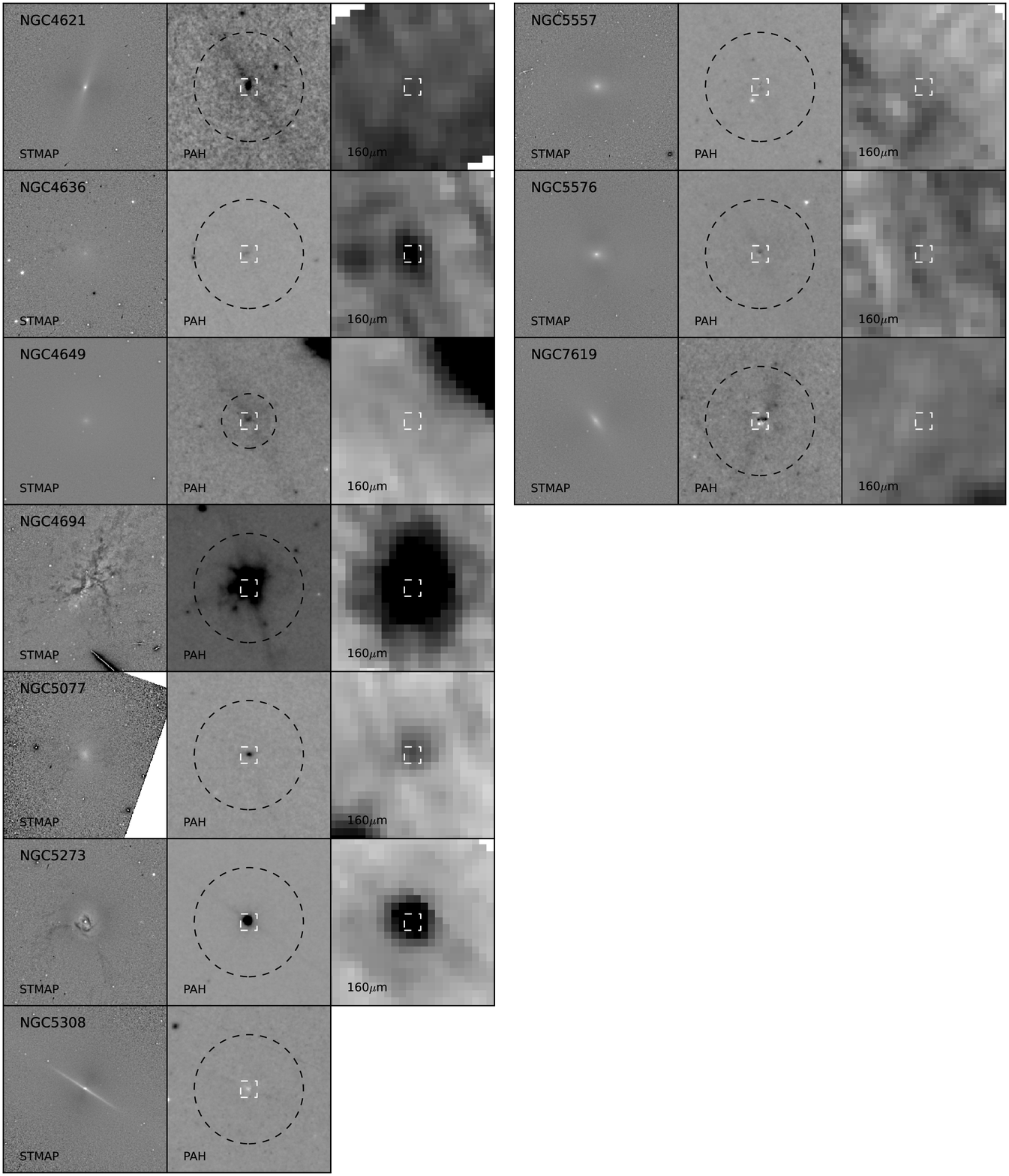} 
\caption{
Continuation of Figure~\ref{fig:multi1}
\label{fig:multi3} 
}
\end{figure*}

\subsection{IRAC Data} \label{sec:irac} 

The PBCD data for these galaxies are sufficiently robust that no additional 
processing was necessary in most cases. The few exceptions were galaxies 
that had relatively few IRAC observations. In these cases we obtained the 
BCD data from the archive and processed them into mosaics with the MOPEX 
software package from the \spitzer\ Science Center. 

We measured aperture photometry in all four IRAC bands with the {\tt phot} task 
in IRAF. We typically adopted an aperture radius of $1'$ because this includes 
nearly all of the observed dust emission at longer wavelengths. In some cases 
we used an $0.5'$ radius due to either nearby, bright objects or the extent of 
the imaging data in one or more bands. In several cases bright, foreground 
stars were in the aperture and we masked them out by replacing those pixel 
values with the median of nearby pixels. The aperture size and photometry for 
each galaxy is listed in Table~\ref{tbl:phot}.  
These measurements include the extended source aperture correction. 

The apertures are smaller than the full extent of the visible emission of 
these galaxies (e.g. $D_{25}$). While these smaller apertures have less noise, 
in some cases they may lead to underestimates of the total emission. 

\begin{deluxetable*}{lrrrrrrrr}
\tablecolumns{9}
\tablewidth{0.0truein}
\tabletypesize{\scriptsize}
\tablecaption{Spitzer Photometry\label{tbl:phot}}
\tablehead{
\colhead{Name} &
\colhead{Aperture} &
\colhead{3.55$\mu$m} &
\colhead{4.49$\mu$m} &
\colhead{5.73$\mu$m} &
\colhead{7.87$\mu$m} &
\colhead{23.68$\mu$m} &
\colhead{71.42$\mu$m} &
\colhead{155.9$\mu$m}
}
\startdata
NGC0315 &    1' &  88.0 &  55.0 &  39.6 &  30.8 &   98.8 (4.0) & 320.3 (33.4) &     475 (58) \\
NGC0821 &  0.5' & 100.9 &  60.6 &  40.2 &  25.1 &    8.8 (0.7) &      $<25.9$ &      $<87.4$ \\
NGC1023 &    1' & 330.0 & 197.8 & 135.9 &  82.7 &   52.3 (2.1) &      $<31.5$ &      $<29.7$ \\
NGC2300 &    1' & 122.6 &  73.0 &  49.4 &  30.2 &     19 (0.8) &      $<16.9$ &      $<66.6$ \\
NGC2768 &    1' & 136.1 &  83.8 &  58.4 &  39.6 &   32.4 (1.3) & 513.2 (51.5) &     486 (61) \\
NGC2787 &    1' & 178.4 & 107.8 &  78.6 &  60.8 &   40.5 (1.6) & 962.9 (96.6) &    833 (103) \\
NGC3226 &  0.5' &  59.6 &  36.0 &  29.1 &  26.9 &     28 (1.2) & 270.2 (28.0)$^p$ &          ... \\
NGC3377 &    1' & 141.1 &  87.4 &  58.5 &  36.1 &   17.2 (0.7) &  55.3 (12.3) &     160 (22) \\
NGC3414 &  0.5' &  98.8 &  59.9 &  42.1 &  30.5 &   26.2 (1.1) &   272 (28.5) &          ... \\
NGC3607 &    1' &   0.0 & 123.5 &   0.0 & 105.4 &     89 (3.6) & 1620.1 (162.3) &   2314 (278) \\
NGC3640 &  0.5' & 138.0 &  84.3 &  56.7 &  35.3 &   19.5 (0.8) &      $<25.6$ &      $<49.8$ \\
NGC3945 &    1' & 133.6 &  83.0 &  57.4 &  34.8 &   34.4 (1.4) & 268.1 (27.4) &          ... \\
NGC3998 &    1' & 199.4 & 127.0 &  96.8 &  74.9 &  149.5 (6.0) & 521.5 (52.5) &     615 (74) \\
NGC4026 &    1' & 151.2 &  97.9 &  67.0 &  40.8 &   20.3 (0.8) &   156 (16.2) &     288 (35) \\
NGC4138 &    1' &  94.1 &  62.1 &  64.6 & 102.8 &  173.9 (7.0) & 2121.2 (212.2) &   3555 (427) \\
NGC4278 &    1' & 201.9 & 122.6 &  86.4 &  62.3 &   46.3 (1.9) & 709.1 (71.4) &     741 (90) \\
NGC4291 &    1' &  84.3 &  51.9 &  34.4 &  20.6 &   21.4 (0.9) &      $<32.9$ &      $<39.1$ \\
NGC4293 &    1' &  78.9 &  61.5 &  72.5 & 114.0 & 528.9 (21.2) & 5193 (519.3) &   6890 (827) \\
NGC4365 &    1' & 191.7 & 116.0 &  81.0 &  49.6 &   31.2 (1.3) &      $<16.0$ &      $<35.3$ \\
NGC4371 &    1' & 107.0 &  63.8 &  43.1 &  27.3 &   15.1 (0.6) &   30.5 (6.3) &      $<29.7$ \\
NGC4382 &    1' & 206.8 & 131.2 &  94.9 &  58.8 &     54 (2.2) &      $<29.0$ &      $<34.4$ \\
NGC4406 &    1' & 206.3 & 122.5 &  83.8 &  51.2 &   40.1 (1.6) &      $<27.3$ &      $<31.6$ \\
NGC4526 &    1' & 310.6 & 191.7 & 170.0 & 224.2 & 284.8 (11.4) & 7599.6 (760.0) &   7049 (846) \\
NGC4550 &  0.5' &  71.9 &  45.0 &  30.2 &  21.4 &    9.6 (0.5) & 166.8 (18.3) &          ... \\
NGC4570 &    1' & 150.7 &  90.5 &  60.0 &  36.9 &   14.5 (0.6) &      $<25.0$ &      $<29.7$ \\
NGC4578 &    1' &  57.1 &  34.9 &  23.3 &  14.3 &    6.9 (0.4) &      $<29.0$ &      $<29.7$ \\
NGC4589 &  0.5' & 103.9 &  64.4 &  44.8 &  29.9 &     14 (0.6) & 204.8 (20.9) &     326 (40) \\
NGC4612 &    1' &  62.3 &  38.7 &  25.8 &  16.4 &    9.6 (0.5) &      $<17.8$ &      $<33.4$ \\
NGC4621 &    1' & 224.6 & 132.1 &  89.8 &  55.5 &   33.2 (1.3) &      $<16.4$ &      $<29.7$ \\
NGC4636 &    1' & 167.2 &  99.4 &  68.2 &  42.4 &     24 (1.0) & 112.8 (16.0) &          ... \\
NGC4649 &  0.5' & 399.7 & 234.7 & 161.6 &  99.1 &   59.4 (2.4) &      $<18.7$ &      $<29.7$ \\
NGC4694 &    1' &  43.1 &  28.2 &  37.4 &  73.2 &  117.5 (4.7) & 1391.9 (139.3) &   2170 (262) \\
NGC5077 &    1' &  93.1 &  56.5 &  38.8 &  25.3 &   20.3 (0.8) & 171.8 (19.2) &     192 (26) \\
NGC5273 &    1' &  39.2 &  27.0 &  24.2 &  27.7 &   95.5 (3.8) &   685 (68.8) &     610 (74) \\
NGC5308 &    1' &  88.5 &  54.1 &  36.7 &  21.9 &    7.6 (0.3) &      $<26.3$ &          ... \\
NGC5557 &    1' &  94.6 &  57.3 &  39.2 &  23.7 &   14.8 (0.6) &      $<13.5$ &      $<33.4$ \\
NGC5576 &    1' & 119.0 &  74.2 &  51.8 &  31.3 &   16.1 (0.7) &      $<19.4$ &      $<32.5$ \\
NGC7619 &    1' &  96.9 &  59.5 &  40.7 &  24.6 &   13.3 (0.5) &      $<18.2$ &      $<29.7$ \\
\enddata
\tablecomments{
Photometry of the galaxies in the sample. Columns are: (1) Galaxy name; 
(2) Aperture radius in arcmin for the 3.6$\mu$m to 70$\mu$m measurements 
(a PSF model was used for the 160$\mu$m photometry, as well as the 
70$\mu$m photometry of NGC3226, which is marked with a $p$ superscript); 
(3-6) Flux in 
IRAC channels 1-4 in mJy; (7-9) Flux in MIPS channels 1-3 in mJy.
Uncertainties include statistical and calibration uncertainties. All 
upper limits are $3\sigma$ upper limits. While NGC4636 is clearly 
detected at 160$\mu$m, we were unable to obtain a good photometric 
measurement.
}
\end{deluxetable*}

\subsection{MIPS Data} \label{sec:mips} 

The \spitzer\ MIPS data were processed further with the MOPEX package. 
In addition to the mosaicing and other processing (e.g.\ median filtering), 
we used the IDL program {\tt BCD column filter.pro} to calculate and subtract 
the median value from each column for each BCD. This process significantly 
diminishes artifacts in the images with minimal loss of flux from the source. 

The fluxes at $24\mu$m and $70\mu$m were measured with aperture photometry 
in the GAIA package and the same aperture sizes employed for the 
IRAC measurements. Aperture corrections were applied to these measurements. 
These corrections were derived from the point spread function (PSF) of bright 
point sources in MIPS 
data. Aperture photometry proved impossible to use for the $160\mu$m fluxes 
because it was difficult to accurately estimate the background. This is because 
the PSF at $160\mu$m is comparable in size to the field of view in many cases. 
There is also the strong possibility of contamination by other sources. 
We therefore extracted the $160\mu$m fluxes with the APEX package in MOPEX, 
which uses a PSF-fitting technique. The standard PSF available with the MOPEX 
package was used for this task. 

We quantified several sources of measurement uncertainty. First, we 
calculated uncertainties for each galaxy in each band from the standard 
deviation of six measurements of the background. For some of 
the $70\mu$m data and most of the $160\mu$m data, this technique may 
not accurately capture the uncertainties as the field of view is too small. 
In these cases we 
set a lower limit on the uncertainty based on the noise in other, similar 
datasets with a larger field of view, such as work by \citet{temi09} and 
\citet{bendo12}, as well as the confusion limit measurement by \citet{frayer09}.
Finally, we included the instrumental flux calibration uncertainties (of 4\%, 
10\%, and 12\% for $24\mu$m, $70\mu$m, and $160\mu$m, respectively). 
Calibration uncertainties dominate the uncertainties for the majority of 
the detections. 

All of the galaxies were easily detected at $24\mu$m and the apertures 
were centered on this emission. Detections at $70\mu$m and $160\mu$m were 
consistent with the same centroid as the $24\mu$m emission. If there was 
not an obvious detection in one of these two bands, then an upper limit 
was measured at the coordinates of the $24\mu$m emission. 
In all cases we quote $3\sigma$ upper limits. Because of the size of our 
aperture, as well as PSF-fitting at $160\mu$m, these upper limits are only 
valid if the dust is centrally concentrated. 

Many of these archival datasets were included in previous studies, in 
particular \citet{temi09} and \citet{bendo12}. Our fluxes generally agree to 
within the uncertainties with \citet{temi09}, who employed a similar set of 
data processing and calibration techniques. The main differences are likely 
because \citet{temi09} employed apertures that were comparable in size to the 
visible extent of the galaxy. Our apertures are smaller, often by a factor of 
several, while at $160\mu$m we employed a PSF-based measurement rather than 
aperture photometry. We have larger differences with the \citet{bendo12} 
measurements and the general sense of these differences is that our 
flux measurements tend to be smaller. Because \citet{bendo12} used 
similar aperture sizes to \citet{temi09}, the difference is likely because 
\citet{bendo12} based their pipeline on an older set of processing and 
calibration tools. Two of the most significant differences we have traced to 
either the presence of a nearby companion (for NGC3226) or clumps of emission 
at large radii \citep[for NGC4406, see also][]{gomez10}. 

As our apertures are smaller than the full extent of the stellar emission 
(although not necessarily the dust emission), our measurements will 
underestimate the flux if the emission is more extended. In most cases these 
differences appear to be at most on the order of a few tens of percent. 
The only significant difference is NGC4526, where our $160\mu$m measurement is 
only 60\% of the value reported by \citet{bendo12}, and about a factor of two 
below the PACS $160\mu$m measurement of \citet{auld13}, which has a much 
larger field of view and consequently better background measurement. We 
discuss the implications of our measurements on the inferred dust masses 
in \S\ref{sec:dustparams}. 

\section{Stellar and Circumstellar Emission} \label{sec:stardust} 


Most of the dust in galaxies is in the interstellar medium. This dust is likely 
composed of a mix of amorphous silicate and graphitic grains, as well as small 
PAH particles. Both stars and PAH particles contribute emission to the $8\mu$m 
IRAC band, so it is important to accurately determine the stellar contribution 
to the $8\mu$m band in order to measure the PAH component. Similarly, both hot, 
circumstellar dust around evolved stars and hot, interstellar dust contribute 
emission to the $24\mu$m MIPS band, so it is important to determine the 
circumstellar contribution to the $24\mu$m band in order to quantify the warm 
component of the diffuse ISM. In the next two subsections we describe how we 
used our sample of galaxies with no evidence for interstellar dust to obtain 
an empirical measurement of the stellar and circumstellar emission 
in the $8\mu$m and $24\mu$m bands.  


\subsection{Stellar Contribution} \label{sec:stars}

IRAC $8\mu$m observations of most galaxies include contributions from 
both stellar photospheres and PAH molecules. In order to study the total flux 
and spatial distribution of PAHs, the conventional practice is to remove the 
stellar contribution by subtraction of a scaled version of the 
$3.6\mu$m flux, where this scale factor is calculated from models of 
stellar photospheres. \citet{helou04} estimated that this value is 
0.232 based on stellar population models from Starburst 99 
\citep{leitherer99} and noted this factor is not very sensitive to 
star formation history or metallicity. They do note that the 
$3.6\mu$m emission may include a hot dust component, based on earlier work 
\citep{bernard94,hunt02}, but estimate that this is a small contribution.
\citet{draine07b} obtained a similar value of 0.260 with the assumption 
that the emission is described by a 5000K blackbody. 

The stellar contribution at $8\mu$m is a small fraction of the total emission 
from the late-type spiral NGC 300 studied by \citet{helou04}. This is also 
true for most of the galaxies in the SINGS sample studied by \citet{draine07b}. 
While approximately half of the galaxies in our sample are similarly expected 
to have substantial dust emission at $8\mu$m, others may have little to no 
dust emission in this band. For these galaxies, an accurate estimate 
of the scale factor specific to early-type galaxy stellar populations is 
necessary to correctly identify and measure PAH emission at $8\mu$m. 

\begin{figure}
\plotone{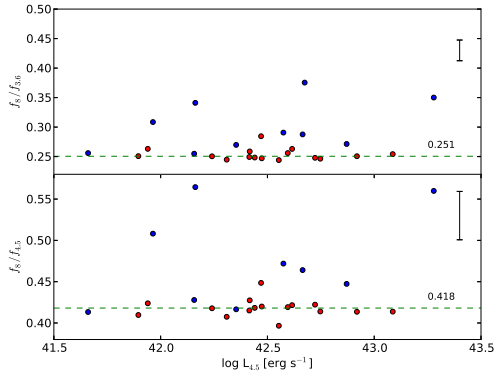} 
\caption{
Ratio of $8\mu$m to $3.6\mu$m flux ({\it top panel}) and $8\mu$m to $4.5\mu$m 
flux ({\it bottom panel}) as a function of $4.5\mu$m luminosity 
for galaxies with the smallest value of this ratio. 
Red points correspond to galaxies without evidence for dust, both based on 
upper limits at $160\mu$m and the absence of dust structure in \hst\ imaging. 
Blue points correspond to galaxies with $160\mu$m detections and dust structure 
in \hst\ imaging. The galaxies with no evidence for dust have median ratios of 
$f_8/f_{4.5} = 0.418 \pm 0.011$ and $f_8/f_{3.6} = 0.251 \pm 0.010$ and we 
adopt these values to 
remove the stellar contribution at $8\mu$m for our early-type galaxy sample. 
The errorbar in each panel represents the average uncertainty in the 
measurement of the ratio. 
Several galaxies with detections at $160\mu$m ({\it blue points}) have 
similar value of this ratio, although some are substantially higher and 
off the upper end of the range shown on this figure. 
\label{fig:iracratios}
}
\end{figure}

We have empirically estimated the appropriate scale factor for early-type 
galaxies with the nominally dust-free galaxies in our sample. 
Figure~\ref{fig:iracratios} shows the ratio of $f_8/f_{3.6}$ and 
$f_8/f_{4.5}$ for the subset of the sample with the smallest values of 
these ratios as a function of $4.5\mu$m luminosity. In both cases all of the 
dust-free galaxies have similar values of these ratios, which correspond to 
$f_8/f_{3.6} = 0.25 \pm 0.01$ and $f_8/f_{4.5} = 0.42 \pm 0.01$. The 
uncertainty quoted for both ratios represents the rms variation of the 
dust-free galaxies. The average photometric uncertainty for the ratio 
measurement is also shown in each panel, although because it includes 
(correlated) absolute calibration uncertainties it is an overestimate
of the true uncertainty in the individual points. 
The measured value for $f_8/f_{3.6}$ is consistent at $1\sigma$ with the 
value of 0.260 chosen by \citet{draine07b} and is $2\sigma$ from the 
value of 0.232 calculated by \citep{helou04}. 

\begin{figure*}
\plotone{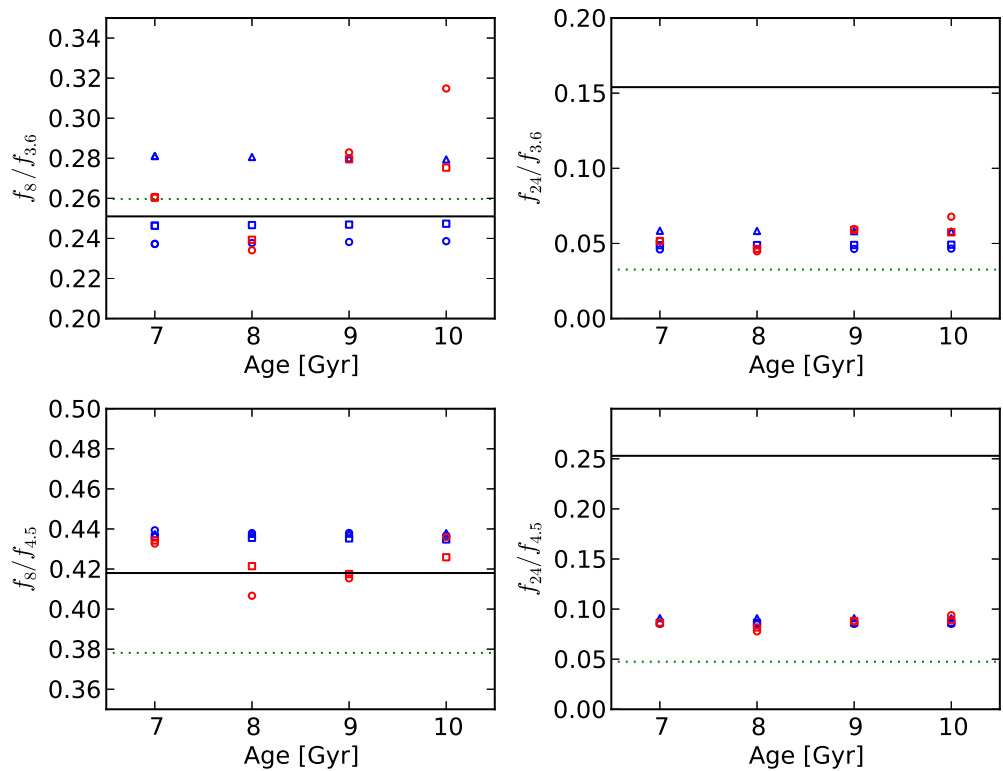} 
\caption{
\spitzer\ IRAC and MIPS band flux ratios for stellar emission from various old 
stellar population models as a function of age (i.e.\ with no ISM or 
circumstellar dust contribution). The flux ratios shown are 
$f_8/f_{3.6}$ ({\it top left}), 
$f_{24}/f_{3.6}$ ({\it top right}), $f_{8}/f_{4.5}$ ({\it bottom left}), 
and $f_{24}/f_{4.5}$ ({\it bottom right}). The model points from 
\citet{maraston05} are for $Z = 0.02$ ({\it blue circles}), $Z = 0.04$ 
({\it blue squares}), and $Z = 0.07$ ({\it blue triangles}). The 
model points from \citet{conroy09} are for $Z = 0.019$ ({\it red circles}) 
and $Z = 0.035$ ({\it red squares}). The empirical ratios ({\it black 
solid line}) and a 5000 K blackbody ({\it dotted line}) are also shown. 
The models are a reasonable representation of the IRAC flux ratios, but 
not the MIPS to IRAC flux ratios, which is evidence for a hot, 
circumstellar dust component. See \S\S\ref{sec:stars} and \ref{sec:hotdust} 
for details. 
\label{fig:scale}
}
\end{figure*}

We have compared these values to the predicted flux ratios from 
stellar population models by \citet{maraston05} and \citet{conroy09} 
to judge how well the models fit these flux ratios, examine the 
expected variation due to stellar population differences, and
estimate the significance of the $24\mu$m detections of apparently 
dust-free galaxies (in \S\ref{sec:hotdust}). As early-type galaxies are 
primarily composed of old stellar population, and this component is 
expected to dominated the stellar emission in the infrared, we only 
examine models with single stellar population ages, but consider metallicites 
from solar to super solar. 
From \citet{maraston05} we calculate the $f_8/f_{3.6}$ and $f_8/f_{4.5}$ 
ratios for metallicities of $Z = 0.02$, $Z = 0.04$, and 
$Z = 0.07$, a Kroupa IMF, and their red horizontal branch prescription. 
From \citet{conroy09} we calculate these ratios for $Z = 0.019$ and 
$Z = 0.035$, a Kroupa IMF, the Padova isochrones, and the Basel library. 

The flux ratios for these models as a function of population age are 
shown in Figure~\ref{fig:scale}. The $f_8/f_{3.6}$ and $f_8/f_{4.5}$ 
ratios from the models agree well with the empirical data 
shown in Figure~\ref{fig:iracratios}. The \citet{maraston05} models
indicate a trend of larger $f_8/f_{3.6}$ ratio at larger metallicities. 
The \citet{conroy09} models appear relatively insensitive to metallicity, 
but do show some increase at larger ages. The $f_8/f_{4.5}$ ratios 
for both sets of models agree well with each other and the data, and 
also appear essentially independent of age and metallicity. Models 
with different IMF prescriptions lead to essentially identical results and 
are not shown. 


\subsection{Circumstellar Dust} \label{sec:hotdust} 

\begin{figure}
\plotone{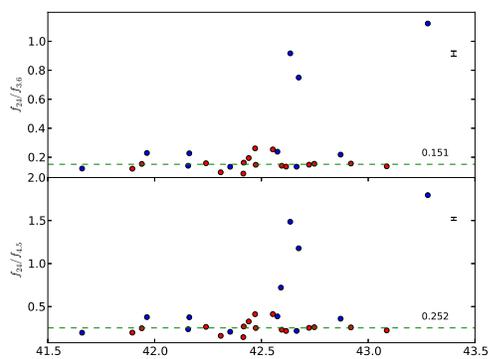} 
\caption{
Same as Figure~\ref{fig:iracratios} for the ratio of $24\mu$m to $3.6\mu$m 
({\it top panel}) and to $4.5\mu$m ({\it bottom panel}). 
Red points correspond to galaxies without evidence for dust, both based on 
upper limits at $160\mu$m and the absence of dust structure in \hst\ imaging. 
Blue points correspond to galaxies with $160\mu$m detections and dust structure 
in \hst\ imaging. The galaxies with no evidence for dust have median ratios of 
$f_{24}/f_{4.5} = 0.252 \pm 0.072$ and $f_{24}/f_{3.6} = 0.151 \pm 0.045$. 
The errorbar in each panel represents the average uncertainty in the 
measurement of the ratio. 
\label{fig:mipsratios} 
}
\end{figure}

While the galaxies without evidence for dust in \hst\ images were also 
undetected at $70\mu$m and $160\mu$m, all are detected at $24\mu$m. 
Figure~\ref{fig:mipsratios} displays a similar plot to 
Figure~\ref{fig:iracratios}, but for $24\mu$m rather than $8\mu$m. This 
figure indicates that the $24\mu$m emission is relatively constant for 
the apparently dust-free galaxies, although the scatter is larger than 
for the $8\mu$m emission. The larger scatter may be due to stellar population 
differences, such as age or metallicity. This was indicated by the work of 
\citet{athey02}, who measured significant intrinsic scatter between the 
mid-infrared and $B-$band luminosities of nine early-type galaxies. 
The ratios and the rms variation we measure are 
$f_{24}/f_{4.5} = 0.253 \pm 0.074$ and $f_{24}/f_{3.6} = 0.154 \pm 0.047$. 
While some of the dusty galaxies also have consistent values of these 
ratios, the ratios for many are much larger and some are not shown on the 
figure. 

These ratios for all galaxies, including the apparently dust-free galaxies, 
are several times larger than expected from stellar photospheric emission 
alone. 
The right panels of Figure~\ref{fig:scale} show the empirical $f_{24}/f_{3.6}$ 
and $f_{24}/f_{4.5}$ ratios compared to the same \citet{maraston05} and 
\citet{conroy09} models described in the previous section. 
In both cases the empirical values are approximately a factor of three higher 
than all of the models. The models are also in good agreement with each other 
and show little evidence for significant variation as a function of stellar 
population age or metallicity. We therefore conclude that the $24\mu$m 
emission is clearly in excess of stellar photospheric emission, yet it does 
scale with the flux of the stellar population. 

The most likely origin of this excess $24\mu$m flux is emission from 
hot dust in the circumstellar envelopes of evolved stars. For a population 
of galaxies with relatively uniform and old stellar populations, 
the fraction of evolved stars with circumstellar dust should be 
an approximately fixed fraction of the stellar mass, which in turn is 
reasonably well represented by the flux at $3.6\mu$m and $4.5\mu$m. 
Models of infrared emission from dust shells around AGB stars by 
\citet{piovan03} include substantial circumstellar emission from $10 - 40\mu$m 
from old stars. This emission includes a local 
maximum that approximately falls within the $24\mu$m band and has been 
identified with silicate emission in oxygen-rich AGB stars \citep{suh02}.  
\citet{bressan07} studied IRS \spitzer\ spectroscopy of many early-type 
galaxies, including a number of the inactive galaxies in our sample 
(NGC 4365, NGC 4371, NGC 4382, NGC 4570, NGC4621), and these data show 
clear evidence for dust emission beyound $8\mu$m that is well fit by a 
dusty silicate circumstellar envelope model \citep{bressan98}. This emission 
spectrum is also well fit by a scaled version of IRS data for stars in the 
globular cluster 47 Tuc studied by \citet{lebzelter06}. 

For galaxies with a substantial amount of cold, interstellar dust ($\geq 
10^6 M_\odot$), such as late-type galaxies, the circumstellar dust 
emission at $24\mu$m is relatively negligible. However, most of the galaxies 
in our sample are substantially more dust poor for their stellar mass.  
It is consequently important to estimate and subtract the circumstellar 
contribution to the $24\mu$m MIPS photometry to obtain an accurate estimate, 
or upper limit, for the cold dust component. 
In the next section, we describe our dust model fits and how we incorporate 
the circumstellar dust emission into the interstellar dust mass estimates and 
upper limits. 


\section{Interstellar Dust} \label{sec:dust} 


Approximately half of these early-type galaxies are detected in the 
MIPS $70\mu$m and $160\mu$m bands, which indicates the presence of 
cold, interstellar dust. In this section we describe how we fit our 
IRAC and MIPS photometry for these galaxies with the dust models of 
\citet{draine07a}, estimate the total dust mass, and compare the 
properties of the dust in early-type galaxies to dust in other morphological 
types. We also derive upper limits to the cold dust mass for 
galaxies that we do not detect at $70\mu$m and $160\mu$m. 

All of the galaxies with dust lanes in the \hst\ structure maps are also 
detected at $8\mu$m. We use the empirical scale factor derived in
\S\ref{sec:stars} to measure the PAH emission in this bandpass and 
examine the spatial distribution of the PAHs. We also estimate the 
fraction of the dust in PAHs with the \citet{draine07a} models for the 
galaxies with suitable data and compare them to other studies. 

\begin{figure*}
\epsscale{.9}
\plotone{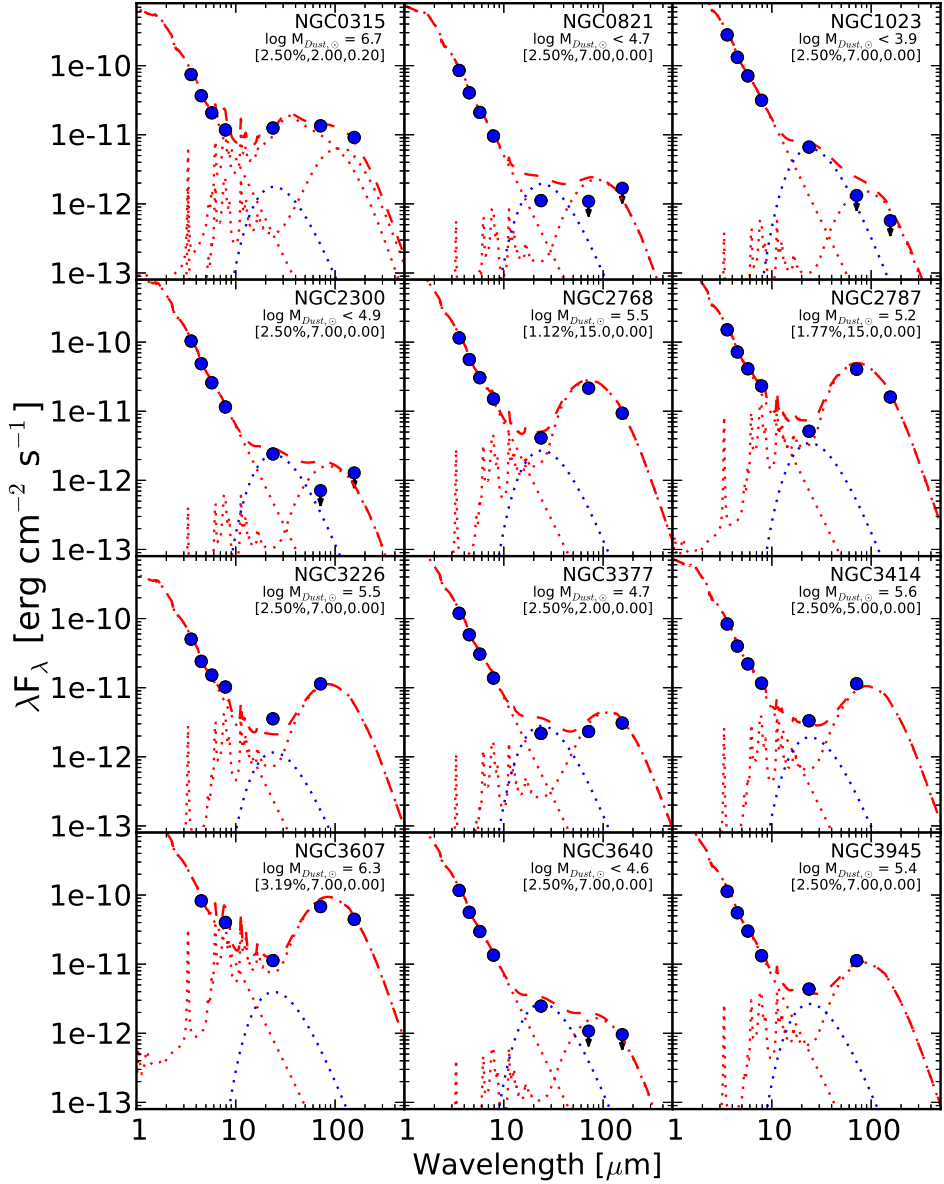} 
\caption{
Infrared spectral energy distributions and model fits. The IRAC and 
MIPS photometric measurements ({\it blue circles} or {\it arrows}) are shown 
with a stellar model from \citet{maraston05}, diffuse and PDR dust models 
from \citet{draine07a} ({\it red, dotted lines}), and 
a simple circumstellar dust model ({\it blue, dotted lines}) that 
corresponds to a fixed fraction of the $4.5\mu$m emission. The sum of 
these components is also shown ({\it red, dashed line}). The dust mass 
estimate, as well as either the best-fit or assumed dust model parameters 
($q_{PAH}$, $U_{min}$, and $\gamma$) are listed below each galaxy's name. 
The dust model fits are described in section \S\ref{sec:model}. 
\label{fig:sed1} 
}
\end{figure*}

\begin{figure*}
\plotone{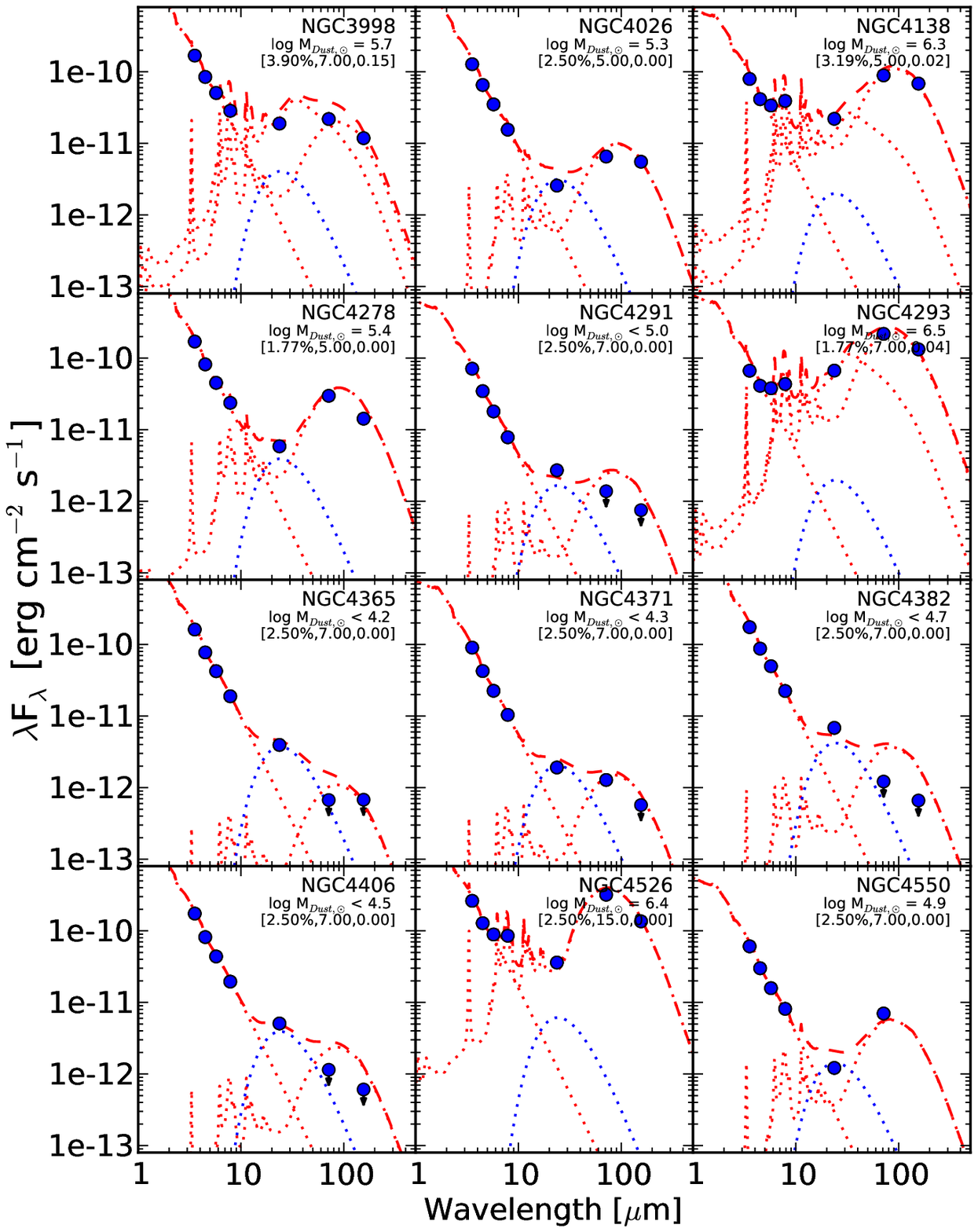} 
\caption{
Continuation of Figure~\ref{fig:sed1}. 
\label{fig:sed2} 
}
\end{figure*}

\begin{figure*}
\plotone{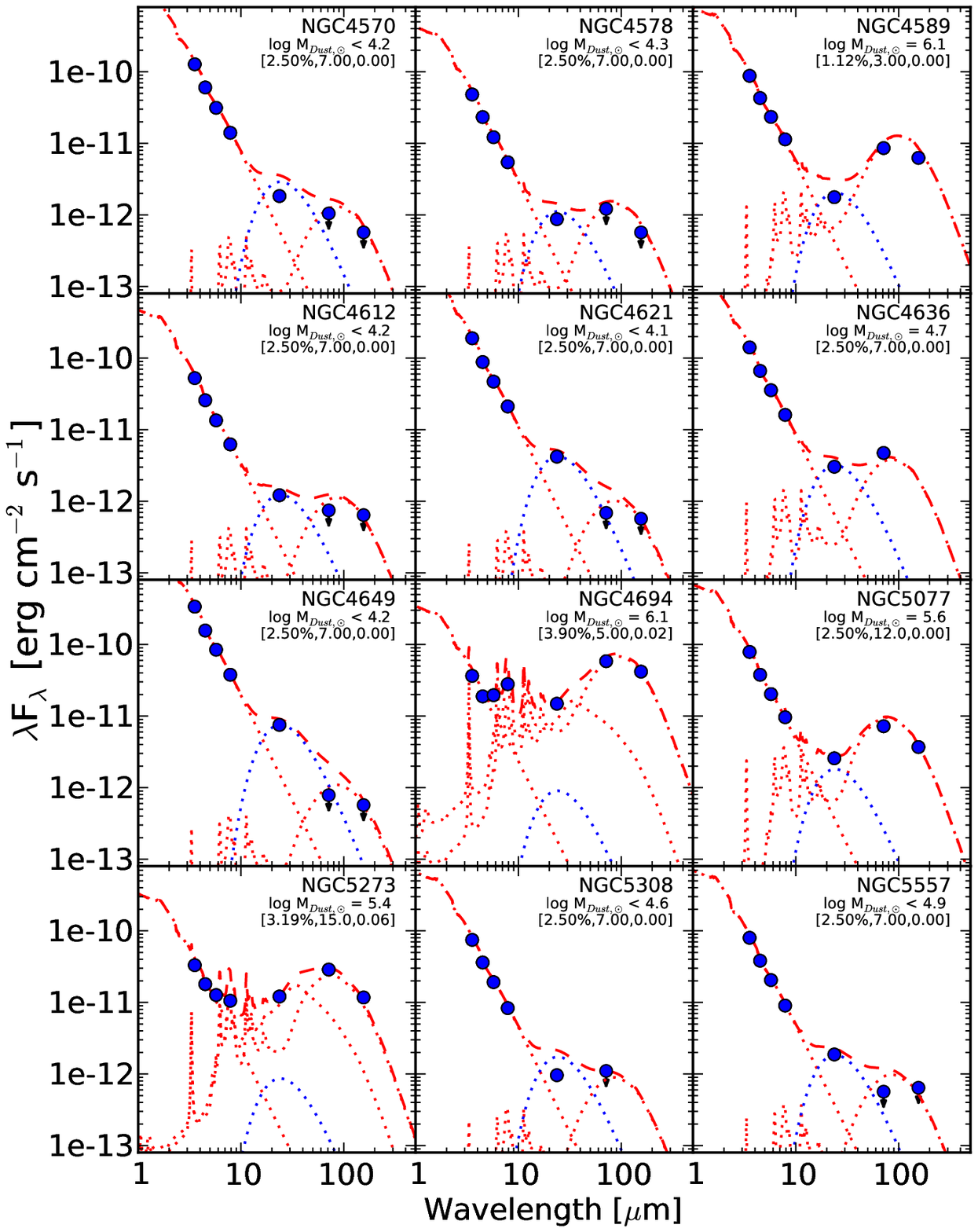} 
\caption{
Continuation of Figure~\ref{fig:sed1}. 
\label{fig:sed3} 
}
\end{figure*}

\begin{figure}
\plotone{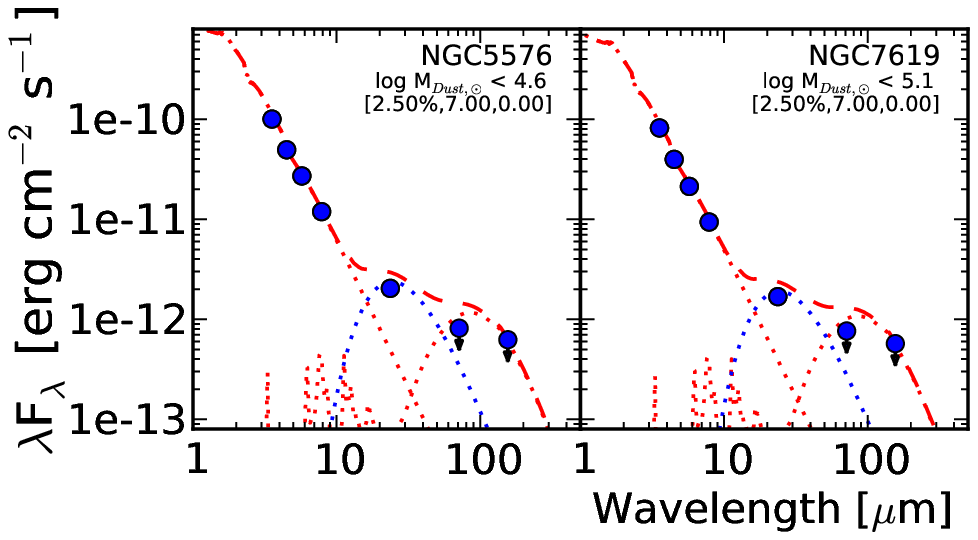} 
\caption{
Continuation of Figure~\ref{fig:sed1}. 
\label{fig:sed4} 
}
\end{figure}

\subsection{Fit to Draine \& Li Models}  \label{sec:model}


Most of the dust in galaxies is in the interstellar medium. This dust 
absorbs energy from starlight, which then re-emits this radiation at 
mid- and far-infrared wavelengths. The emission spectrum of this 
dust can be used to estimate the total mass in dust, as well as other 
properties, given a model for the typical grain size distribution 
and composition. These models are used to characterize the absorption and 
re-emission of the Interstellar Radiation Field (ISRF) as a function of 
wavelength and may consequently be used to derive the dust mass from the 
observed luminosity. 


We employ the models of \citet{draine07a} to characterize the properties 
of the dust in these galaxies and derive dust masses. These models 
assume that dust is comprised of a mixture of carbonaceous and silicate 
grains, and varying amounts of PAH particles. The main observational 
constraints on the grain size distribution, relative mixture and opacities of 
carbonaceous and silicate grains, and the properties of the PAH particles 
are the wavelength-dependent extinction observed in the Milky Way 
\citep{mathis83}, long-wavelength observations of the Milky Way 
\citep{finkbeiner99}, and studies of the properties of PAH emission 
from the Milky Way and nearby galaxies \citep[e.g.][]{smith07}. 


The mid- to far-infrared emission from dust grains depends on the incident 
ISRF, and thus a characterization of the distribution of starlight intensities 
is also necessary to infer the properties of the dust. 
Integrated measurements of entire galaxies will typically include emission 
from dust that has been heated by a wide range of different values of the 
local ISRF. While some studies perform detailed radiative 
transfer calculations to estimate the local value of the ISRF for individual 
dust grains \citep[e.g.][]{popescu00,gordon01,piovan06}, this approach 
typically relies on many assumptions and 
may not be necessary to obtain reasonable estimates of the dust mass and 
other dust properties. \citet{draine07a} propose a simple parametrization 
of the ISRF where a fraction $\gamma$ of the dust mass is exposed to 
high-intensity radiation, such as may occur near the sites of star formation, 
and the remainder is exposed to the average ISRF. The dust exposed to a wide 
range of ISRF intensities, which \citet{draine07a} refer to as dust 
associated with Photodissociation Regions (or simply PDR dust) 
is exposed to a power-law distribution of intensities from $U_{min}$ to 
$U_{max}$, where $U$ is a dimensionless scale factor such that 
$U = 1$ for the local ISRF derived by \citet{mathis83}, $U_{min}$ is 
the minimum value of the ISRF, $U_{max}$ is the maximum value, and the 
power-law index is $\alpha$. 
The dust exposed to the average ISRF is exposed to a radiation field of 
intensity $U_{min}$ and is referred to as diffuse dust. The diffuse 
dust is expected to be relatively cold and dominate the emission in the 
$70\mu$m and $160\mu$m bands, while if present the PDR dust may be important 
at shorter wavelengths, most notably the $24\mu$m MIPS band. \citet{draine07b} 
find that the SEDs of galaxies in the SINGS sample are satisfactorily 
reproduced with $U_{max} = 10^6$ and a power-law index of $\alpha=2$. We 
fix these two model parameters at those values for this study as well. The ISRF 
for the dust models is consequently just characterized by $U_{min}$ and 
$\gamma$. As early-type galaxies have relatively little star formation, we 
do not expect large values of $\gamma$; however, $\gamma$ may not be 
completely negligible as some early-type galaxies have modest amounts of 
star formation. The PDR component may also represent dust near the active 
nucleus in the small number of moderate luminosity AGN in this sample. 


Very small dust grains and PAHs are subject to single-photon heating 
to high temperatures and thus emit at much shorter wavelengths than 
larger grains \citep{sellgren84}. PAHs are also very likely to be 
the cause of the emission features observed at 3.3, 6.2, 7.7, 8.6, and 
11.3 $\mu$m in the Milky Way and most galaxies, and in particular 
the $7.7\mu$m feature appears to dominate the nonstellar emission 
in the IRAC $8\mu$m band. This emission has been identified 
with C-C stretch and C-H bending modes, primarily from ionized PAHs. 
The \citet{draine07a} models
include PAHs as small as 20 carbon atoms and note that PAHs with
$<1000$ carbon atoms dominate the emission at wavelengths $\lambda < 20\mu$m.
They present seven different Milky Way dust models that have different size 
distributions of very small grains and different PAH fractions $q_{PAH}$ 
of 0.5\% to 4.6\%, where $q_{PAH} = 4.6$\% is a reasonable match to the 
Milky Way. Because each of these seven dust models has a different 
PAH fraction, $q_{PAH}$ uniquely describes the dust model. The 
dust emission from a given galaxy is consequently characterized by 
$q_{PAH}$, $U_{min}$, $\gamma$, and the total dust mass $M_{dust}$. 
In the next subsection we describe how we derive these four parameters. 

\subsubsection{Flux Ratios} 


The $8\mu$m, $24\mu$m, $70\mu$m, and $160\mu$m bands may be dominated by dust 
emission. We use the data from these four bands to determine the four 
parameters of the \citet{draine07a} model: $q_{PAH}$, $U_{min}$, $\gamma$, and 
$M_{dust}$. Our approach uses the graphical procedure described in 
\citet{draine07a} to estimate the best-fit dust model. This procedure employs 
three flux ratios to determine the model that is the best fit to the spectral 
shape. The total luminosity is used to estimate the dust mass in the context 
of the best-fit model. The ratios 
defined in that paper are: 
\begin{equation}
P_{7.9} = \frac{\langle \nu f_\nu^{ns} \rangle_{7.9}}{\langle \nu f_\nu \rangle_{71} + \langle \nu f_\nu \rangle_{160}}
\end{equation} 
\begin{equation}
P_{24} = \frac{\langle \nu f_\nu^{ns} \rangle_{24}}{\langle \nu f_\nu \rangle_{71} + \langle \nu f_\nu \rangle_{160}}
\end{equation} 
\begin{equation}
R_{71} = \frac{\langle \nu f_\nu \rangle_{71}}{\langle \nu f_\nu \rangle_{160}}
\end{equation} 
The superscript ``ns'' refers to the nonstellar contribution to that bandpass. 
For the $8\mu$m band, this is the value after subtraction of the empirical 
stellar contribution with the procedure described in \S\ref{sec:stars}. 
For the $24\mu$m band, this is the value after the subtraction of the 
empirical stellar and circumstellar dust contribution to the $24\mu$m band 
as described in \S\ref{sec:hotdust}. The measured values of these ratios are 
listed in Table~\ref{tbl:model}. 

Not all of the galaxies with detections at $8\mu$m, $70\mu$m, and $160\mu$m 
have a value for $P_{7.9}$ in Table~\ref{tbl:model} because in three cases 
(NGC3377, NGC4026, and NGC5077) the flux of the non-stellar component at 
$8\mu$m is consistent with the uncertainty in the empirical scale factor 
and our aperture measurements, even though weak PAH emission is apparent in 
Figures~\ref{fig:multi1} -- \ref{fig:multi3}. For a similar reason, no 
$P_{24}$ value is quoted for seven galaxies (the previous three cases, as well 
as NGC2768, NGC2787, NGC4278, and NGC4589) because there is more 
substantial variation in the empirical scale factor that corrects for 
the contribution from (mostly) hot, circumstellar dust in this passband. 

\begin{deluxetable}{lccclrlr}
\tablecolumns{8}
\tablewidth{0.0truein}
\tabletypesize{\scriptsize}
\tablecaption{Dust Model Fits\label{tbl:model}}
\tablehead{
\colhead{Name} &
\colhead{P7.9} &
\colhead{P24} &
\colhead{R71} &
\colhead{$q_{PAH}$ [\%]} &
\colhead{$U_{min}$} &
\colhead{$\gamma$} &
\colhead{log $M_{dust}$}
}
\startdata
NGC0315 &   0.13 &   0.48 &   1.47 &  2.50 &    2.0 &    0.2 &    6.7 \\
NGC0821 &    ... &    ... &    ... &  2.50 &    7.0 &    0.0 & $<$4.7 \\
NGC1023 &    ... &    ... &    ... &  2.50 &    7.0 &    0.0 & $<$3.9 \\
NGC2300 &    ... &    ... &    ... &  2.50 &    7.0 &    0.0 & $<$4.9 \\
NGC2768 &   0.06 &    ... &   2.31 &  1.12 &   15.0 &    0.0 &    5.5 \\
NGC2787 &   0.11 &    ... &   2.52 &  1.77 &   15.0 &    0.0 &    5.2 \\
NGC3226 &    ... &    ... &    ... &  2.50 &    7.0 &    0.0 &    5.5 \\
NGC3377 &    ... &    ... &   0.75 &  2.50 &    2.0 &    0.0 &    4.7 \\
NGC3414 &    ... &    ... &    ... &  2.50 &    5.0 &    0.0 &    5.6 \\
NGC3607 &   0.18 &   0.07 &   1.53 &  3.19 &    7.0 &    0.0 &    6.3 \\
NGC3640 &    ... &    ... &    ... &  2.50 &    7.0 &    0.0 & $<$4.6 \\
NGC3945 &    ... &    ... &    ... &  2.50 &    7.0 &    0.0 &    5.4 \\
NGC3998 &   0.25 &   0.44 &   1.85 &  3.90 &    7.0 &   0.15 &    5.7 \\
NGC4026 &    ... &    ... &   1.18 &  2.50 &    5.0 &    0.0 &    5.3 \\
NGC4138 &   0.19 &   0.13 &   1.30 &  3.19 &    5.0 &   0.02 &    6.3 \\
NGC4278 &   0.10 &    ... &   2.09 &  1.77 &    5.0 &    0.0 &    5.4 \\
NGC4291 &    ... &    ... &    ... &  2.50 &    7.0 &    0.0 & $<$5.0 \\
NGC4293 &   0.10 &   0.19 &   1.65 &  1.77 &    7.0 &   0.04 &    6.5 \\
NGC4365 &    ... &    ... &    ... &  2.50 &    7.0 &    0.0 & $<$4.2 \\
NGC4371 &    ... &    ... &    ... &  2.50 &    7.0 &    0.0 & $<$4.3 \\
NGC4382 &    ... &    ... &    ... &  2.50 &    7.0 &    0.0 & $<$4.7 \\
NGC4406 &    ... &    ... &    ... &  2.50 &    7.0 &    0.0 & $<$4.5 \\
NGC4526 &   0.12 &   0.07 &   2.35 &  2.50 &   15.0 &    0.0 &    6.4 \\
NGC4550 &    ... &    ... &    ... &  2.50 &    7.0 &    0.0 &    4.9 \\
NGC4570 &    ... &    ... &    ... &  2.50 &    7.0 &    0.0 & $<$4.2 \\
NGC4578 &    ... &    ... &    ... &  2.50 &    7.0 &    0.0 & $<$4.3 \\
NGC4589 &   0.08 &    ... &   1.37 &  1.12 &    3.0 &    0.0 &    6.1 \\
NGC4612 &    ... &    ... &    ... &  2.50 &    7.0 &    0.0 & $<$4.2 \\
NGC4621 &    ... &    ... &    ... &  2.50 &    7.0 &    0.0 & $<$4.1 \\
NGC4636 &    ... &    ... &    ... &  2.50 &    7.0 &    0.0 &    4.7 \\
NGC4649 &    ... &    ... &    ... &  2.50 &    7.0 &    0.0 & $<$4.2 \\
NGC4694 &   0.23 &   0.14 &   1.40 &  3.90 &    5.0 &   0.02 &    6.1 \\
NGC5077 &    ... &    ... &   1.95 &  2.50 &   12.0 &    0.0 &    5.7 \\
NGC5273 &   0.15 &   0.28 &   2.45 &  3.19 &   15.0 &   0.06 &    5.4 \\
NGC5308 &    ... &    ... &    ... &  2.50 &    7.0 &    0.0 & $<$4.6 \\
NGC5557 &    ... &    ... &    ... &  2.50 &    7.0 &    0.0 & $<$4.9 \\
NGC5576 &    ... &    ... &    ... &  2.50 &    7.0 &    0.0 & $<$4.6 \\
NGC7619 &    ... &    ... &    ... &  2.50 &    7.0 &    0.0 & $<$5.1 \\
\enddata
\tablecomments{
Dust model results of the galaxies in the sample. Columns are: (1) Galaxy 
name; (2) P7.9; (3) P24; (4) R71; (5) $q_{PAH}$; (6) $U_{min}$; 
(7) $\gamma$; (8) log of $M_{dust}$ in solar masses.
Galaxies with measurements of P7.9, P24 and/or R71 were fit with the 
\citet{draine07a} models. The remaining galaxies were fit with a fixed set of 
parameters: ($q_{PAH}$, $U_{min}$, $\gamma$) = (2.50\%, 7.0, 0.0). Dust mass 
upper limits are for this fixed model and the $3\sigma$ upper limits at 
70$\mu$m and 160$\mu$m. More details are provided in \S5.1.
}
\end{deluxetable}


The ratio $P_{7.9}$ is effectively a measure of the fraction of the dust 
emission that is radiated by PAHs, and thus provides the strongest 
constraint on the dust model and the $q_{PAH}$ parameter. 
\citet{draine07a} note that because the $8\mu$m emission is almost 
exclusively due to single photon heating, $P_{7.9}$ is not very sensitive 
to the starlight intensity. The exception is that if the 
PAH fraction is very small, exposure to a more intense ISRF (larger $\gamma$) 
will lead to enhanced PAH emission that may produce values of $P_{7.9}$ that 
are comparable to models with larger $q_{PAH}$ and smaller values of $\gamma$. 
$P_{7.9}$ is less sensitive to $\gamma$ for larger values of $q_{PAH}$ 
because these galaxies exhibit substantial PAH emission even when 
$\gamma = 0$. 


The value of $\gamma$ is mostly determined by the ratio $P_{24}$. This 
ratio is sensitive to the fraction of the dust emission near $24\mu$m 
and is dominated by hot, PDR dust ($\gamma>0$) if a hot component is present. 
According to \citet{draine07a}, when larger grains are heated by starlight 
intensities above $U \sim 20$, they will contribute to the $24\mu$m 
emission. In contrast, dust that is heated by starlight intensities of 
$U \sim 0.1 - 10$ is sufficiently cold that it emits most of its luminosity 
at longer wavelengths, such as the $70\mu$m and $160\mu$m bands. 


The coldest dust component is diffusely distributed in the ISM and heated 
by the average ISRF of the galaxy, $U_{min}$. The ratio $R_{71}$ is sensitive 
to $U_{min}$ because the intensity of the ISRF is correlated with the 
characteristic temperature of the cold component, and this affects the ratio 
of the flux in the $70\mu$m and $160\mu$m bands. For small values of $\gamma$ 
(on order a few percent or less), the ratio $R_{71}$ is a very sensitive to 
$U_{min}$ because the $70\mu$m flux has a negligible contribution from PDR 
dust. For larger values of $\gamma$, the PDR dust contribution to the 
$70\mu$m emission is more important. In galaxies with substantial star 
formation, the values of both $R_{71}$ and $P_{24}$ are larger and 
$R_{71}$ is less sensitive to $U_{min}$. 



\subsubsection{Dust Model Parameters} \label{sec:dustparams} 

\begin{figure*}
\plotone{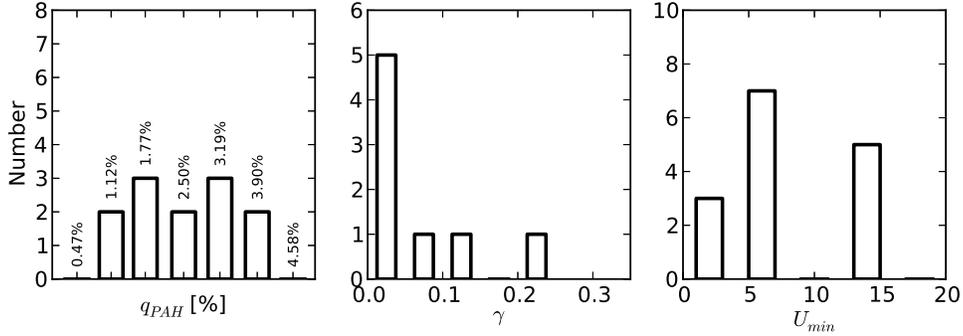} 
\caption{
Distribution of the dust model parameters $q_{PAH}$ ({\it left}), 
$\gamma$ ({\it middle}), and $U_{min}$ ({\it right}) in galaxies 
with sufficient detections to constrain one or more of these quantities. 
Note that NGC3998 has $\gamma = 1$ and consequently does not fall within 
the range shown in the middle panel. 
\label{fig:modelhist}
}
\end{figure*}

There are \nmips\ galaxies in our sample with sufficient data to constrain 
all three \citet{draine07a} model parameters. For these galaxies we follow the 
graphical procedure recommended by \citet{draine07a} and first estimate 
$q_{PAH}$ from the values of $P_{7.9}$ and $R_{71}$, and then use the values 
of $P_{24}$ and $R_{71}$ to estimate $\gamma$ and $U_{min}$. 
Four additional galaxies have just $P_{7.9}$ and $R_{71}$, which is sufficient 
to constrain the dust model. A histogram 
of $q_{PAH}$ for these 11 galaxies is shown in the first panel of 
Figure~\ref{fig:modelhist}. The median value of $q_{PAH}$ is $2.50$\% for 
these galaxies. The second panel shows the distribution of $\gamma$ for the 
subset of \nmips\ galaxies with sufficient coverage of the diffuse emission.  
The median value value of $\gamma$ for these eight is $0.03$. Finally, there 
are three galaxies with just $R_{71}$. For these galaxies we assume $q_{PAH} 
= 2.50$\% and $\gamma = 0$ (to provide a more conservative constraint on the 
dust mass than $\gamma = 0.03$) and then estimate $U_{min}$. The median value 
of $U_{min}$ is 7.0. 

\begin{figure}
\plotone{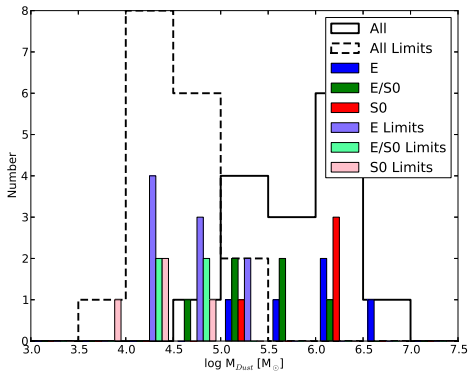}
\caption{
Distribution of dust masses and upper limits as a function of morphological 
type. 
\label{fig:mdusttype} 
}
\end{figure}

\begin{figure}
\plotone{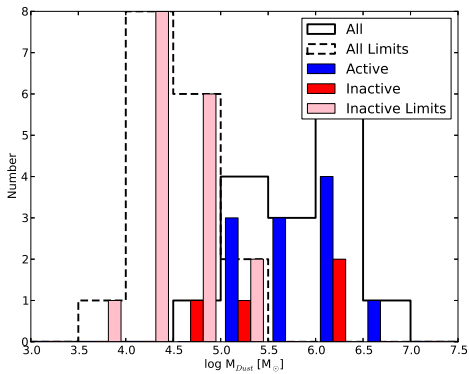}
\caption{
Distribution of dust masses and upper limits as a function of nuclear activity. 
\label{fig:mdustactive} 
}
\end{figure}

With the dust model fixed, the dust mass is: 
\begin{equation}
M_{dust} = \frac{\Psi}{\langle U \rangle} \left( \langle \nu f\nu \rangle_{24} + \langle \nu f \nu\rangle_{71} + \langle \nu f\nu \rangle_{160} \right) D^2. 
\end{equation} 
The quantity $\Psi$ is uniquely determined by the three parameters $q_{PAH}$, 
$\gamma$, and $U_{min}$ that describe the dust model, the quantity 
$\langle U \rangle$ accounts for starlight heating in both PDRs and the 
diffuse ISM, and $D$ is the distance. 
Dust mass histograms are shown for different morphological types in 
Figure~\ref{fig:mdusttype} and for active and inactive galaxies in 
Figure~\ref{fig:mdustactive}. 

The distributions of $q_{PAH}$, $\gamma$, and $U_{min}$ are generally different 
from the distributions found by \citet{draine07b} for the SINGS sample. 
The $q_{PAH}$ values are generally lower, with the median of $q_{PAH} = 
2.5$\% below the median value of $3.4\%$ found by \citet{draine07b} 
for the 61 galaxies in the SINGS sample with similar IRAC and MIPS data. 
The $8\mu$m PAH feature is mostly due to ionized PAHs \citep{joblin94}, so the 
weakness of this feature may indicate 
a greater fraction of neutral PAHs, rather than a smaller PAH fraction. 
\citet{kaneda05} proposed this explanation based on their study 
of IRS spectra for four elliptical galaxies, which showed the short-wavelength 
PAH features were weak or absent, but the longer wavelength features at 
$11.3\mu$m and $12.7\mu$m were prominent. Their sample includes 
NGC4589, for which our best model fit yields $q_{PAH} = 1.12$\%, yet the longer 
wavelength PAH features are clearly present in the IRS spectrum. 
\citet{kaneda05} suggest the relative absence of UV radiation in ellipticals 
could produce physical conditions amenable to primarily neutral PAHs. Such 
conditions are also indicated by the spectral energy distribution of the 
bulge of M31 by \citet{groves12}. 

Other potential explanations include the preferential destruction of small 
particles, such as due to sputtering in hot gas or an AGN, and low gas-phase 
metallicities. The dominance of larger PAHs over smaller ones, which could be 
a consequence of sputtering \citep{schutte93,tsai95,micelotta10}, would weaken 
the $8\mu$m feature. The $q_{PAH}$ value is also observed to decline in 
galaxies with low gas-phase oxygen abundance. \citet{draine07b} find that 
the median PAH fraction is $1.0$\% for galaxies with gas-phase oxygen 
abundances below $12 + {\rm log}_{10} {\rm O/H} < 8.1$ (it is also low for a 
small fraction of higher metallicity galaxies). Even though our sample 
is expected to possess solar to super-solar stellar abundances, the cold 
gas abundance may be quite low if it is the remant of an accreted, low-mass  
satellite. Finally, relatively luminous 
AGN have also been implicated in the destruction of PAHs \citep{lutz08} and 
all but three of the galaxies with detections in all seven bands are 
AGN, although they are low luminosity AGN. \citet{smith07} found that AGN may 
modify the relative distribution of intensities of various PAH features, and 
in particular suppress those from $5-8\mu$m. However, low-luminosity AGN do 
not appear to negatively effect the PAH abundance, as \citet{draine07b} found 
that the median $q_{PAH}$ fraction was actually slightly higher (3.8\%) for 
AGN than the median of their full sample, as well the median of 28 galaxies 
with HII nuclei (3.1\%). 

Nearly all of the galaxies have $\gamma < 0.1$ and the median value is 
$\gamma = 0.03$. This is consistent with the near-absence of star formation 
in these galaxies. The small values of $\gamma$ also mean that the 
model fits and dust masses are more reliable, as in this regime the 
parameter $R_{71}$ provides a better estimate of $U_{min}$. The median 
$U_{min}$ value for this sample is $7.0$, which is several times higher 
than the SINGS sample median of 1.5. The median for the subsample of 
SINGS galaxies with SCUBA data is in better agreement with our data. 
In the case of SINGS galaxies with SCUBA observations, the cold and PDR dust 
components are better separated, and thus $U_{min}$ is better measured for 
those galaxies. We may be able to measure larger values of $U_{min}$ without 
longer-wavelength sub-mm observations because our galaxies have lower 
values of $\gamma$. Early-type galaxies may also have more intense ISRFs than 
spiral galaxies due to the presence of hot, evolved stars and higher stellar 
densities. This is also suggested by their higher dust temperatures relative 
to spirals \citep{bendo03,skibba11,smith12,auld13}. 
Emission from hot, evolved stars contributes to and may dominate the 
LINER emission spectra from some galaxies \citep{sarzi10}. Longer wavelength 
observations would help further constrain the value of $U_{min}$, while 
far-infrared line ratios would help to constrain the spectral shape of the 
ionizing photons \citep[e.g.][]{malhotra00}. 


An important caveat that affects the interpretation of the dust model 
parameters is that uncertainties, or intrinsic variations, in the empirical 
scale factors derived in Section~\ref{sec:stars} increase the uncertainties 
in the dust model parameters. As noted previously, the data shown in 
Figures~\ref{fig:iracratios} and \ref{fig:mipsratios} exhibits evidence for 
intrinsic variation in these quantities, and some variation is expected due 
to stellar population differences. This is particularly relevant for 
$P_{7.9}$ and $P_{24}$, which largely determine $q_{PAH}$ and $\gamma$, 
and so these parameters are not as well constrained as $U_{min}$. In 
practice, as $\gamma$ is expected to be small due to the general absence 
of star formation in these galaxies, the biggest potential uncertainty 
lies with $q_{PAH}$. 



The measured dust masses for our sample are $M_{dust} \sim 10^{5-6.5}$ \msun, 
which are two or three orders of magnitude lower than the dust masses for 
spiral galaxies of comparable size, but similar to other early-type galaxies 
\citep{goudfrooij95,draine07b,smith12}. 
The dust masses that result from this procedure are fairly robust, even though 
they are only based on measurements out to $160\mu$m. This is true 
even if the model parameters $q_{PAH}$, $\gamma$, and $U_{min}$ are not well 
constrained by the data because $q_{PAH}$ has a negligible impact on the dust 
mass estimate and $\gamma = 0$ is a good approximation for most galaxies. 
For the full range of models presented in \citet{draine07a}, we also note 
that the value of $\Psi$ that is set by the best-fit model only varies 
from $\sim 0.044 - 0.066$ g (erg s$^{-1}$)$^{-1}$. This indicates that even if 
the dust model were not constrained at all by the data and one simply adopted 
the average of this range, there would only be a $\sim 20$\% uncertainty in the 
dust mass within the context of these models. An additional uncertainty 
is the ISRF because the absence of flux measurements at wavelengths longer 
than $160\mu$m do not constrain a potentially substantial mass of cold dust. 
However, the available data can constrain the ISRF quite well because these 
galaxies have little star formation. Another estimate of the uncertainties 
comes from the overlap between several galaxies in our study 
and previous investigations with ISO \citep{xilouris04a}, \spitzer\ 
\citep{temi07a}, and longer-wavelength Herschel data \citep{smith12,auld13}. 
The dust masses derived by those previous studies assumed a modified 
black body and mostly agree with our estimates to within a factor 
of two. 
The noteworthy exception is NGC3945, for which \citet{smith12} derive
over an order of magnitude larger dust mass. The difference could be ascribed 
to the fact that we use a smaller aperture, although our $70\mu$m flux      
is within 5\% of the IRAS $60\mu$m value. However, due to the large 
uncertainty in this IRAS measurement it is still possible that we are missing 
extended emission from dust in this object. We may be able to put tighter 
constraints on the dust mass for this object with future analysis of Herschel
photometry. We also note we were unable to obtain a good measurement of NGC3945 
at $160\mu$m. We do retain NGC3945 in our sample. 
Another comparison was performed by \citet{gordon10}, who found that dust 
masses estimated from $\leq160\mu$m 
data can differ by $10-36$\% relative to dust mass estimates that include 
Herschel measurements at $250\mu$m and $350\mu$m, although this study was 
based only on the observations of the Large Magellanic Cloud and therefore 
does not include uncertainties due to differences in grain size and ISRF 
intensity distributions. Nevertheless, this study and a recent one by 
\citet{aniano12} of two, nearby spirals that also compared \spitzer\ 
and Herschel provides good evidence against substantial masses of cold dust. 

A final uncertainty is that the physical dust model that works so well 
for the Milky Way may not be appropriate for early-type galaxies, that is the 
size and composition are different because of different formation and 
processing histories. For example, sputtering can reduce the fraction of 
small grains relative to the initial grain size distribution. The generally 
smaller values of $q_{PAH}$ in early-type galaxies relative to spirals may 
indicate that this is the case, although there is also allowance for this 
difference in the \citet{draine07a} models. The study of four ellipticals by 
\citet{kaneda05} also provides some evidence that the shape of the radiation 
field is softer in ellipticals than in the Milky Way, which may not be 
adequately compensated by an adjustment in $U_{min}$. Provided all early-type 
galaxies are different in the same manner, then the relative dust masses we 
have derived here should nevertheless be robust. The true uncertainties in the 
dust masses are likely dominated by these model details and are difficult to 
quantify. Under the assumption that the dust model is a reasonably good match 
to early-type galaxies, we estimate that the uncertainties in the dust masses 
are on order a factor of two or three due to a combination of aperture 
differences, calibration uncertainties, and differences (or uncertainties) in 
the dust models. 

\subsubsection{Upper Limits} 


Approximately half of the galaxies in our sample are not detected at $70\mu$m 
and $160\mu$m. These non-detections nevertheless provide very useful 
constraints on the total dust mass in these galaxies because of the 
extraordinary sensitivity of the \spitzer\ MIPS instrument. To calculate upper 
limits on the dust masses in these galaxies, we employ a dust model with 
$q_{PAH} = 2.50$\%, $\gamma = 0$, and $U_{min} = 7.0$. The values of $q_{PAH}$ 
and $U_{min}$ are representative of the galaxies with the best data 
(see Figure~\ref{fig:modelhist}). We chose to use $\gamma = 0$, rather than 
the median value of $\gamma = 0.03$, because this leads to a more conservative 
constraint on the dust mass. 

We use this dust model to calculate the largest value of the dust mass that is 
consistent with the $3\sigma$ upper limits at $70\mu$m and $160\mu$m. These 
upper limits are listed in Table~\ref{tbl:model} and shown in 
Figures~\ref{fig:mdusttype} and \ref{fig:mdustactive}. The figures clearly 
show that the upper limits on the dust masses are approximately an order 
of magnitude lower than the measured dust masses. 

All of these upper limits are for galaxies that show no evidence of dust lanes 
within 100s of parsecs of their nuclei based on our earlier \hst\ study 
\citep[][and also shown in the leftmost panels of Figures~\ref{fig:multi1} -- 
\ref{fig:multi3}]{lopes07}.  
While the structure map technique is insensitive to uniformly
distributed or very diffuse dust, the \spitzer\ data place very strong 
upper limits on the mass of such dust. The PAH images are also valuable 
in this respect, as we describe in the next subsection. 

\subsection{PAHs} 





As a second search for diffuse, extended dust that may not have been present in 
our \hst\ study, we created PAH images for all of the galaxies. These images 
are shown in the middle panels of Figure~\ref{fig:multi1} -- \ref{fig:multi3}.  
The PAH images are the $8\mu$m image minus a scaled version of the 
$4.5\mu$m image, where the scale factor accounts for the relative stellar 
emission in the two images. The calculation of the scale 
factor is described in \S\ref{sec:stars}. The pair of images were convolved 
to the same PSF prior to subtraction. 
All of the galaxies with dust lanes in the \hst\ structure maps also have clear
PAH emission, with the exception of NGC4636, which has only marginally-detected 
dust lanes in the \hst\ data. Conversely, none of the galaxies with no evidence 
for dust at visible wavelengths exhibit clear evidence of PAH emission. 
A number of the PAH images of the galaxies exhibit artifacts at their 
centers. These are consistent with artifacts of the stellar continuum 
subtraction, which could be caused by uncertainties in the scale factor, 
the PSF shape, and centroid errors. 


There is a good correspondence between the morphology of the dust lanes in 
the \hst\ data and the distribution of PAH emission for the small subset of 
galaxies where the \hst\ data are of sufficient quality and the dust lanes 
sufficiently extended. Some particularly striking examples are 
NGC4138 and NGC4293. In a few cases the structure maps show substantial 
asymmetries in the dust distribution that are not 
mirrored in the PAH images. The most likely explanation 
is that the dust lanes viewed in absorption are substantially 
less prominent on the side of the galaxy that is tilted away 
from our perspective \citep[e.g.][]{hubble43}. 

\subsection{Distribution of Dust Masses} \label{sec:dustdist} 


The dust mass distribution shown in Figure~\ref{fig:mdusttype} shows that most 
detections have $10^{5-6}$ \msun\ of dust. There are only three detections 
below $10^{5}$ \msun\ (NGC3377, NGC4550, and NGC4626), while there are many 
upper limits that extend an order of magnitude lower and suggest a relative 
absence of galaxies with $<10^5$ \msun\ of dust. For example, if there were as 
many galaxies with dust masses in the range $10^{4-5}$ \msun\ as in the range 
$10^{5-6}$, we would expect ten galaxies in this lower range and that we 
would detect half of them, yet only one is detected. This apparent dichotomy is 
also supported by the similar dichotomy in the detection of dust lanes in 
\hst\ images of these galaxies (see Table~1), as we do not expect the 
\spitzer\ and \hst\ images to have comparable dust mass sensitivity. 

One possible explanation is a difference in morphological type. 
Figure~\ref{fig:mdusttype} splits the detections and upper limits by 
morphology into ellipticals, S0s, and E/S0s. We see no evidence that true 
ellipticals or true S0 galaxies are more or less likely to have interstellar 
dust. The sensitivity of the datasets for each morphological type are also 
comparable. The upper limits on the dust mass in undetected ellipticals and 
lenticulars both span the full range of upper limits from approximately 
$10^{4-5}$ \msun. We used the Astronomical Survival Analysis (ASURV) programs 
from \citet{feigelson85}, as implemented in the {\tt IRAF STSDAS STATISTICS} 
package, to compare the elliptical and lenticular subsamples. We ran five 
different tests through the {\tt twosampt} task and these indicated that 
the distributions are consistent. These similar distributions are in contrast 
to the results of \citet{smith12} with the \herschel\ Reference Survey, who 
detected dust in 24\% of ellipticals and 62\% of S0s with SPIRE. 
This difference in detection rate between the two morphological types could 
be due to their larger sample of 62 early-type galaxies. Another potential 
explanation is that \citet{cortese12} and found that the dust to stellar 
mass ratio is lower for ellipticals than lenticulars. If our ellipticals 
were more massive than our lenticulars, this would also explain their 
similar dust masses. 

\begin{figure}
\plotone{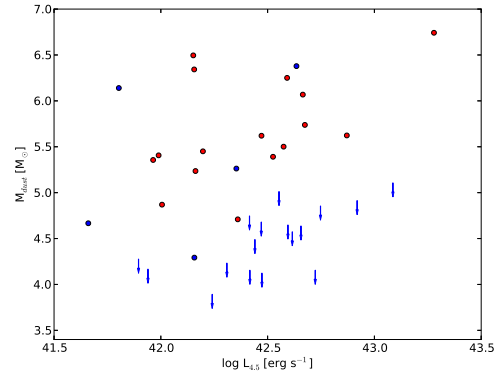}
\caption{
Dust mass $M_{dust}$ vs.\ luminosity in at $4.5\mu$m $L_{4.5}$ for galaxies 
with good dust mass estimates ({\it solid points}) and upper limits 
({\it arrows}). Active galaxies are marked with red symbols and inactive
galaxies with blue symbols. 
The upper limits are based on the fiducial dust model 
described in \S5.1.3 and the $3\sigma$ upper limits on the flux at 
$70\mu$m and $160\mu$m. 
\label{fig:mdustlum} 
}
\end{figure}


We investigated if there is a correlation between dust detections or 
upper limits and cluster membership. 
Environment may correlate with the absence of dust because the hot atmospheres 
of clusters may lead to an increase in the rate of dust destruction, although 
many cluster early-type galaxies do maintain their own atmospheres 
\citep[e.g.][]{sun05}. 
With regard to environment, clusters of galaxies have substantially fewer 
gas-rich dwarf galaxies and sufficiently high relative velocities that mergers 
are more rare. We searched for differences as a function of environment, and 
whether or not environment could explain the dust ditchotomy, with the 14 
Virgo members and 27 nonmembers in our sample and detect dust in 18/27 
(66.6\%) nonmembers, compared to 5/14 (45.4\%) Virgo members. This difference 
suggests a slightly larger incidence of dust in galaxies outside of Virgo, 
although it is not statistically significant. \citet{smith12} performed a 
similar analysis for the early-type galaxies in the Herschel Reference Survey 
and found a similar trend, but that it was also not statistically significant. 
We also investigated whether or not membership influenced the detection of 
dust in the inactive sample. We detect dust in 2 of the 11 inactive Virgo 
members, compared to three out of 11 of the inactive galaxies outside of 
Virgo. There is thus no evidence that membership in the Virgo cluster is 
correlated with a lower incidence of dust in inactive galaxies. 

We also compared the dust detection rate with stellar luminosity, which 
correlates well with stellar mass. More massive galaxies generally 
have higher ISM temperatures, which will lead to faster dust destruction 
rates. This naively suggests an anticorrelation between dust mass and stellar 
luminosity, although no such anticorrelation is present in our data. 
In Figure~\ref{fig:mdustlum} we show the dust masses and upper limits as a 
function of $L_{4.5}$, a proxy for stellar mass. While the minimum detected 
dust mass is a function of stellar luminosity, this is because more luminous 
galaxies are more rare and thus at larger distances. 

The dichotomy between the dust mass estimates and upper limits thus suggests 
that there are distinct populations of dusty and non-dusty early-type galaxies, 
rather than a continuum of dust masses. To explore this possibility further, 
we resampled our data many times in order to determine if the apparent 
dichotomy is due to the sensitivity distribution of our data, rather than an 
intrinsically nonuniform distribution of dust masses. To perform these 
calculations, we estimated the minimum flux sensitivity of the $160\mu$m data 
for each galaxy. We then generated a uniform distribution in log $M_{dust}$, 
randomly assigned dust masses from this distribution to galaxies in our 
sample, calculated the flux at $160\mu$m based on a randomly selected galaxy's 
distance for the dust model used to estimate upper limits, and determined if 
this flux would correspond to a detection or upper limit based on the 
sensitivity of the same galaxy's dataset. We generated 1000 realizations of 
our dataset and found that the sensitivity and distance distribution of our 
dataset does not produce the observed dichotomy between detections and upper 
limits for an intrinsically uniform distribution in log $M_{dust}$ 
approximately $\sim 95$\% of the time. 

While this is formally marginally significant, an important caveat to this 
statistical analysis is that it does not include variations due to the 
best-fit dust model, which may add a factor of two uncertainty to the dust 
mass, the possibility that we may systematically underestimate the dust mass 
by tens of percent in some cases, particularly for nearby galaxies with 
extended emission, and potential correlations with morphology, stellar mass, 
and environment. We plan to reexamine this potential dichotomy with a larger 
sample. 


\subsection{Correlation with AGN} \label{sec:agn} 


The distribution of dust masses is strikingly different 
for active and inactive galaxies. Galaxies classified as active, that is either 
LINERs or Seyferts according to \citet{ho97b}, are all detected and have dust 
masses that range from $M = 10^{5-6.5}$ \msun. In contrast, only four galaxies 
classified as inactive are detected and the remaining 16 have upper limits 
that range from $M \sim 10^{4-5}$ \msun. This result is similar to the strong 
correlation between activity and the presence of dust structure within the 
central kpc in our earlier \hst\ study \citep{lopes07}, yet these new 
observations demonstrate that the galaxies without dust lanes have at least 
one to two orders of magnitude less dust than those 
with detections. As in the previous section, we used the ASURV programs from 
\citet{feigelson85} to compare the dust mass distributions of the active and 
inactive samples. All five tests available through the {\tt twosampt} task 
returned zero probability that the dust masses were drawn from the same 
distribution. 

There has been substantial discussion in the literature of the extent to 
which the emission lines characteristic of LINERs are actually powered by 
accretion onto supermassive black holes. Other proposed candidates 
are fast shocks, interaction with the hot phase of the ISM, and photoionization 
by UV emission from old stellar populations \citep[see][for a review]{ho08}. 
\citet{sarzi10} performed a detailed analysis of emission-line ratios for 
many early-type galaxies and concluded that the emission-line luminosities 
of only a minority of LINERs are dominated by accretion. Nevertheless, AGN have 
been confirmed in the majority of LINERs based on measurements of radio cores, 
X-ray cores, and the surface brightness profiles of various emission lines 
relative to the stellar continuum 
\citet{filho04,nagar05,filho06,flohic06,ho08,sarzi10}. Some fraction of the 
cold ISM in the majority of these dusty early-type galaxies is therefore 
fueling their central, supermassive black holes, even if their emission line 
luminosities are not dominated by nuclear accretion. We hypothesize that the 
correspondence between active galaxies and dust in our sample is 
simply due to the presence of a sufficient quantity of cold material,  
some of which has accreted onto the black hole, and some of 
which is ionized and produces the visible-wavelength emission lines, rather 
than a causal relationship between the presence of dust and AGN. 


\section{Origin of the Dust}  \label{sec:origin} 


The main motivation for this study was to use dust mass estimates for 
early-type galaxies to constrain the origin of their dust. Our estimates 
demonstrate that dusty early-type galaxies have $10^{5-6.5}$ \msun\ of dust, 
while the remainder have at least one order of magnitude less. 
Our sample is representative of the early-type galaxy population as a 
whole because the original selection of the sample was from the 
magnitude-limited Palomar Survey of \citet{ho95}. In \citet{lopes07}, 
we selected an equal number of active and inactive galaxies from 
all of the early-type galaxies in the Palomar survey that had 
visible-wavelength \hst\ images. Half of this sample were active 
galaxies and half were not, which is equivalent to the relative frequency of 
active and inactive galaxies in the early-type galaxy population. 
\citet{lopes07} then found that all of the galaxies classified as 
active had dust lanes within the central kpc, whereas only 25\% of the 
inactive galaxies had them. While we have more inactive galaxies than 
active galaxies in this sample, we still find that all active galaxies 
have interstellar dust and find a similarly small fraction 
(5/22 or 23\%) of the inactive galaxies have dust. We therefore 
conclude that $\sim 60$\% of all early-type galaxies have dust, 
including all of those classified as active, and $\sim40$\% do not. 


The presence of dust in any early-type galaxy has long been a puzzle because 
thermal sputtering of dust grains in the hot ($T>10^{6}$ K) gas that comprises 
the bulk of their ISM should destroy dust on the timescale 
$\tau_{dust} \sim 2 \times 10^{4}$ yr \citep{draine79a}. These galaxies 
clearly do have evolved stars that produce dust, as indicated by the strong 
evidence for hot, circumstellar dust, and this dust must be ejected into the 
ISM at an approximately constant rate due to the age of their stellar 
populations. The expected rate of mass loss, the dust-to-gas ratio of this 
material, and the dust destruction timescale, can be used to estimate the 
steady-state dust mass of these galaxies. 
As noted in the Introduction, for reasonable estimates of the mass-loss 
rate, dust-to-gas ratio, and dust destruction time in hot gas, the 
steady-state dust mass 
should be $10 - 100 (\tau_{dust}/2 \times 10^4 {\rm yr})$ \msun, that is
none of the galaxies should have been detected. 


The dust destruction timescale could be longer if the temperature of the hot 
phase is closer to $10^5$ K, or if some of the gas and dust cools rapidly and 
settles into a cooler phase \citep{mathews03}, although the dust destruction 
timescale needs to be many orders of magnitude longer to produce dust masses of 
$10^{5}$ \msun\ and above. 
A further complication is that this calculation predicts the expected 
steady-state dust mass for all early-type galaxies. As early-type 
galaxies form a fairly homogeneous population, all early-type galaxies should 
have similar dust production rates and destruction timescales. The internal 
origin model is consequently inconsistent with the order of magnitude 
differences in dust mass for similar galaxies, such as is illustrated by 
Figure~\ref{fig:mdustlum}. 

The many galaxies with upper limits on the dust mass provide an upper limit 
on the typical dust destruction timescale in the hot ISM through an inversion 
of the argument presented above. For typical early-type galaxies, 
the most stringent upper limits on the dust mass are approximately 
$M_{dust} < 10^{4-4.5}$. For a continual dust production rate of 
$0.1 - 1 \times 0.005 = 5 - 50 \times 10^{-4}$ \msunyr, 
the upper limit on the dust destruction timescale is: 
\begin{equation}
\tau_{dust} < 2 \times 10^7 \left( \frac{M_{dust,lim}}{10^4}\right) \left(\frac{0.1}{\dot{M}_{gas}}\right) \left(\frac{0.005}{\epsilon_{dg}}\right)  yr
\end{equation} 
where $M_{dust,lim}$ is the upper limit on the dust mass, $\dot{M}_{gas}$ 
is the gas mass loss rate of the evolved stellar population, and
$\epsilon_{dg}$ is the dust to gas ratio. 
This upper limit is comparable to the value of $\tau_{dust} < 46 \pm 25$ Myr 
derived by \citet{clemens10} with a similar argument, as well as more 
recent work by \citet{smith12}. 
Both values are consistent with, although several orders of magnitude greater 
than, the calculations of \citet{draine79a}, and inconsistent with an internal 
origin for the one to two orders of magnitude more dust present in 
approximately half of the sample. As noted in the introduction, a number of 
previous studies have also pointed out the difficulties with a purely 
internal origin \citep{lopes07,rowlands12}. In particular, the recent study by 
\citet{smith12} reached a similar conclusion with a similar analysis of 
62 early-type galaxies observed by \herschel. 


The alternate, and favored, hypothesis for the origin of the dust is the 
accretion of gas-rich satellites \citep[e.g.][]{tran01,verdoeskleijn05}. The 
morphology of the dust lanes observed in \hst\ images indicate that 
many of them are chaotic and clumpy amd this appears inconsistent with 
steadily accumulation from stellar mass loss \citep{vandokkum95,lopes07}. 
Furthermore, when smooth dust disks are present, they do not always share the 
major axis of the host galaxy. Kinematic observations of atomic 
\citep[e.g.]{bertola84,bertola92,sarzi06,morganti06} and molecular 
\citep{combes07,davis11} gas in early-type galaxies also indicate that the gas 
kinematics are often not aligned with the kinematics of the stars, which 
further suggests an external origin. 

The key challenges to the external origin hypothesis are that the merger rate 
of gas-rich satellites is low and the dust destruction timescale is short, yet 
dust is present in slightly more than half of all early-type galaxies. 
That is, the external origin model must satisfy the constraint: 
\begin{equation}
f_{dust} = \Re_{merg} \tau_{dust}
\end{equation}
where $f_{dust} = 0.6$ is the fraction of the population with dust above 
some minimum mass, $\Re_{merg}$ is the merger rate of gas-rich satellites 
that could supply sufficient dust, and $\tau_{dust}$ is the dust destruction 
time. Dwarf galaxies with $10^{5-6}$ \msun\ of dust (comparable in 
size to the Magellanic clouds) would correspond to $1:300 - 1:100$ mass-ratio 
mergers for the galaxies in this sample. Based on the equations provided by 
\citet{stewart09}, we calculate that the cumulative merger rate for mass 
ratios from equal mass mergers to $1:300$ mergers is $\Re_{merg} = 0.07 - 0.2$ 
Gyr$^{-1}$, which is more than two orders of magnitude higher than our upper 
limit of $\tau_{dust} < 0.02$ Gyr and more than four orders of magnitude 
greater than the theoretical estimate \citep{draine79a}. These values predict 
$f_{dust} < 0.0014 - 0.004$, or a discrepancy of $150 - 430$. 

To evaluate the significance of this discrepancy, we estimate the uncertainty 
in both quantites. First, the merger rate we have adopted may be an 
overestimate because the theoretical calculations predict the merger rates 
for all galaxies as a function of stellar mass, yet not all satellites 
galaxies may have sufficient cold gas. It is also possible that the 
theoretical estimates underestimate the true merger rates. Observational 
evidence for an underestimate was presented by \citet{lotz11}, who found that 
the observed merger rates are approximately a factor of five higher than the 
predictions of \citet{stewart09} and other recent, theoretical analyses. 
\citet{lotz11} posit that the disagreement between theory and observation 
could be because the visibility timescale for the mergers has been 
overestimated, or the theoretical rate has been underestimated. While the 
merger visibility timescale is arguably more uncertain that the merger rates 
from simulations, this comparison indicates the merger rates are unlikely more 
than a factor of a few larger than the theoretical estimates. There is 
consequently insufficient uncertainty in the merger rates to resolve this 
discrepancy. 

The other potential resolution is that the dust destruction timescale is 
actually substantially higher than both the theoretical estimate of 
\citet{draine79a} and our empirical upper limit. While this 
requires an increase of about two orders of magnitude relative to the 
upper limit, the dust destruction timescale is much more uncertain than the 
merger rate. One reference value is the dust destruction timescale for the 
Milky Way $\tau_{dust,MW} = 0.4$ Gyr, which is dominated by the impact of 
supernova shocks on the cold, neutral medium 
\citep{barlow77,draine79b,dwek80,jones94}. 
As the supernova rate in early-type galaxies is several times lower than in the 
Milky Way \citep[e.g.][]{li11}, the dust destruction timescale may be 
several times larger in their cold ISM and plausibly on the order of a Gyr, 
provided the accreted dust does not mix with the hot ISM. 
This may be the case if 
a much larger dust destruction timescale, combined with the highest estimated 
merger rates, predicts a dusty fraction that is only a factor of a few below 
what is needed to reproduce the observations. This is plausibly within the 
uncertainties of these order of magnitude estimates, particularly if dust 
destruction is very inefficient in the cold ISM of early-type galaxies and 
the cold ISM does not mix efficiently with the hot phase.
However, it is also plausible that the lifetime will be less because 
there is less cold ISM. A more precise estimate could be obtained with 
an analysis of the fraction of the supernova energy that impacts the 
cold ISM. 



While it may be possible for purely external accretion to work, we instead 
propose that the solution is a hybrid of internal production and external 
accretion. When early-type galaxies accrete a gas-rich satellite, they accrete 
substantial amounts of both cold gas and dust. 
While dust will be destroyed by a 
combination of supernova shocks and sputtering, this destruction could be 
balanced by continued dust grain growth in the accreted cold ISM. While this 
process is not constrained by observations of early-type galaxies, many studies 
of the Milky Way have concluded that $>90$\% of dust formation by mass occurs 
in the cold ISM \citep{draine79b,dwek80,mckee89,draine09}, rather than 
from stellar sources such as mass loss and supernova alone 
\citep{barlow10,matsuura11,dunne11}. 
The Milky Way has $M_{gas} = 
5 \times 10^9$ \msun\ of cold gas that produces $M_{dust} = 2.5 \times 10^7$ 
\msun\ per dust destruction timescale, which corresponds to a steady-state dust 
mass of $\epsilon_{dg} M_{gas}$. While the dust destruction timescale is 
likely different from the Milky Way value, the steady-state dust mass is 
likely to be within an order of magnitude of that predicted by the product of 
the Milky Way dust-to-gas ratio and the observed cold gas supply. 

Continued dust production in the accreted cold gas removes the main weakness 
of the internal origin hypothesis because the mass of neutral gas in 
early-type galaxies varies by orders of magnitude \citep[e.g.][]{welch10}, 
even though their stellar populations are quite similar. 
The hybrid solution also removes the main weakness of the external origin 
hypothesis because dust destruction is at least partially balanced by growth 
in the cold gas. As a result, the timescale that is important to explain the 
demographics of dust is the cold gas depletion timescale, rather than the 
destruction timescale of individual grains. As the depletion timescale for 
the cold gas appears to be on the order of Gyr, largely due to star formation, 
it is sufficient to maintain substantial quantites of dust in approximately 
half of all early-type galaxies. 

This hybrid solution leads to several testable predictions. First, the 
cold gas in early-type galaxies should be lower metallicity than the old 
stellar population, which may be accessible with gas-phase abundance 
measurements. Second, the dusty early-type galaxies should have cold gas to 
dust ratios comparable to the accreted satellite population, unless the 
cold gas and dust are destroyed at different rates. Finally, there should 
be more evidence of mergers within the last few Gyr in the dusty early-type 
galaxies than those that appear dust free. 

Both external accretion and our hybrid solution also constrain the 
lifetime of the low-luminosity AGN in the dusty early-type galaxies. If the 
only source of the cold gas and dust that fuels the supermassive black hole 
is external accretion, and these accretion events only occur every few Gyr, 
the high incidence of AGN implies that these galaxies remain (weakly) active 
for several Gyr as well, while the inactive galaxies remain inactive for 
comparable timescales. The mass of dust, and the two orders of magnitude more 
cold gas in the ISM, is more than sufficient to maintain accretion rates of 
$\leq 0.01$ \msun\ yr$^{-1}$ for this time, as well as fuel the modest 
circumnuclear star formation neccesary to produce the nuclear stellar disks 
seen in many dust-free early-type galaxies \citep[e.g.][]{lopes07}. 



\section{Summary} \label{sec:summary} 

We have analyzed the dust in a representative sample of
\ntot\ early-type galaxies and used these data to constrain the origin 
of their dust. This analysis is largely based on archival \spitzer\ 
IRAC and MIPS observations, although these galaxies also 
have \hst\ observations that reveal the presence or absence of 
dust lanes in the central kpc, as well as high SNR, visible-wavelength 
spectroscopy that provides uniform classifications of nuclear activity. 

The IRAC and MIPS observations detect emission from dust from every galaxy 
that exhibits dust lanes seen in absorption in \hst\ observations. This dust 
emission is 
clearly due to both PAH particles, as detected with IRAC $8\mu$m 
observations, and cold, interstellar dust, as detected by MIPS observations 
at $70\mu$m and $160\mu$m. Conversely, galaxies that show no evidence for 
dust lanes in \hst\ observations also do not exhibit dust emission at longer 
wavelengths. The absence of dust emission indicates that galaxies without 
clumpy dust lanes within 100s of parsecs of their centers do not have a 
substantial quantity of uniform, diffusely-distributed dust that would 
have been difficult to detect with contrast enhancement techniques. 

The galaxies with no evidence for interstellar dust are a valuable sample to 
estimate the stellar spectral energy distribution for old stellar populations. 
We use these data to estimate the ratio of the stellar emission at $8\mu$m to 
that at $3.6\mu$m and $4.5\mu$m. We then use these flux ratios to scale 
and subtract the stellar contribution from the $8\mu$m band for the galaxies 
with interstellar dust and obtain a good estimate of the emission from PAHs 
in this bandpass. These flux ratios are in reasonable agreement with 
spectral synthesis models of old stellar populations. 

We also used the galaxies without interstellar dust to estimate the 
ratio of the flux in the $24\mu$m MIPS band to the shorter-wavelength, 
photospheric emission at $3.6\mu$m and $4.5\mu$m. The ratio of the 
$24\mu$m emission to these two shorter wavelength bands is fairly 
constant across galaxies of a wide range of luminosity, yet the ratio 
exceeds that expected from photospheric emission by approximately a constant 
factor of four. We conclude that the discrepancy is because emission 
from hot dust in the circumstellar envelopes of evolved stars makes a 
substantial contribution to the $24\mu$m band. We use this ratio to remove the 
circumstellar dust contribution from the early-type galaxies that have dust 
in order to study the dust properties of the interstellar medium. 

A number of these galaxies have sufficient diffuse dust emission that 
we can determine one or more of the dust model parameters introduced by 
\citet{draine07a}. These are the PAH fraction $q_{PAH}$, the fraction 
of the dust heated by a strong radiation field $\gamma$, and the minimum 
starlight intensity $U_{min}$. The best-fit model and the far-infrared 
luminosity then determine the dust mass. This analysis indicates 
that early-type galaxies are usually well fit by the same dust models 
that work well for spiral galaxies. We do find evidence that early-type 
galaxies have slightly lower values of $q_{PAH}$, which may be due to the 
presence of primarily neutral PAHs and/or reflect low ISM metallicity. These 
galaxies also have lower values of $\gamma$ and higher values of $U_{min}$ 
than typical spirals. 

The dust masses of these early-type galaxies are at least several orders of 
magnitude lower than the dust masses of spirals. The dust mass of a typical 
early-type galaxy detected in the far-infrared is $10^{5-6.5}$ \msun; however, 
many galaxies are undetected in the long-wavelength MIPS bands 
and the upper limits on the dust masses in these galaxies range up to an 
order of magnitude lower, or $10^{4-5}$. As our sample was designed to be 
representative of the entire early-type galaxy population, our results 
indicate that approximately $60$\% of typical early-type galaxies have 
$\geq 10^5$ \msun\ of interstellar dust. 

There is no correlation between the presence of dust and galaxy morphology. 
Our sample contains approximately equal numbers of ellipticals and lenticular 
galaxies and the same fraction of each morphological type has interstellar 
dust. We also see no strong evidence for a correlation between the presence 
of dust and environment. 
There is a strong correlation between the detection of dust and the presence 
of visible-wavelength emission lines, which are usually consistent with some 
form of nuclear activity. We detect dust emission from all of the galaxies 
with nuclear activity, while only approximately 25\% of the galaxies classified 
as inactive have detectable dust emission. Based on the relatively weak 
nuclear emission in these galaxies, we conclude that the dust, and the 
larger reservoir of cold gas associated with it, is a necessary but not 
sufficient condition for accretion onto the center, supermassive black hole. 

We use our dust mass measurements and the demographics of dust in early-type 
galaxies to demonstrate that most of the dust mass can not originate from 
evolved stars. The main argument is that there are orders of magnitude 
variations in the dust mass between galaxies with the same morphological type 
and stellar mass. We also use these data to rule out an exclusively external 
origin for the dust. An external origin requires that the product of the 
merger-rate of gas-rich galaxies and the dust destruction timescale equals the 
fraction of dusty early-type galaxies. While many observations of dust lane 
morphology and gas kinematics provide strong evidence for an external origin, 
the observed dust masses require mergers of gas-rich galaxies that are 
at least an order of magnitude too rare to explain the presence of dust in the 
majority of early-type galaxies. The dust destruction timescale would need 
to be nearly an order of magnitude greater than the Milky Way value of 
$4 \times 10^{8}$ years, rather than the expectation that it is $10^{4-5}$ 
years due to sputtering in hot gas, in order for purely external accretion to 
be viable. 

We propose instead that the external accretion of cold gas provides an 
environment for the continual growth of dust for much longer than the nominal 
destruction timescale of individual grains. If dust grains continue to grow 
in mass in the externally-accreted cold medium, the reduction in the 
steady-state dust 
mass should scale with the much longer depletion timescale of the cold gas, 
rather than the typical destruction timescale for individual 
grains. In fact, continual dust grain growth in this cold gas seems 
inevitable, provided the conditions are similar to those that lead to 
dust grain growth in the Milky Way. The longer gas depletion timescale of 
several gigayears is consistent with the merger rates of gas-rich satellites, 
the morphology and kinematics of the cold dust and gas, and the distribution 
of dust masses. In this scenario, the lifetime for the low-luminosity AGN 
present in most dusty, early-type galaxies is also comparable to this several 
gigayear timescale. 

\acknowledgments

We thank Ramiro Sim{\~o}es Lopes for his assistance with the early stages of 
this project. We are grateful to Karin Sandstrom and the referee for many 
comments on the manuscript, as well as to Alison Crocker, Adam Leroy, and 
Craig Sarazin for helpful discussions. PM is grateful for support from the 
sabbatical visitor program at the North American ALMA Science Center (NAASC) 
at NRAO and the hospitality of both the NAASC and the University of Virginia 
while this work was completed.
This work is based on archival data obtained with the Spitzer Space Telescope, 
which is operated by the Jet Propulsion Laboratory, California Institute of 
Technology under a contract with NASA. Support for this work was provided by 
an award issued by JPL/Caltech.

{\it Facilities:} \facility{HST}, \facility{SPITZER}.


\end{document}